\documentclass[10pt,aps,prb,superscriptaddress,longbibliography,twocolumn]{revtex4-2}

\usepackage{setspace}
\usepackage{lmodern}

\usepackage[latin9]{inputenc}
\usepackage{amsmath}
\usepackage{amssymb}
\usepackage{graphicx}
\usepackage{epstopdf}
\usepackage{color}
\usepackage{enumerate}
\usepackage{array}
\usepackage{bm}
\usepackage{tabularx}
\usepackage{mathtools} 
\usepackage{multirow}
\usepackage{hyperref}
\usepackage[all]{hypcap}

\hypersetup{
    colorlinks=true,
    linkcolor=blue,
		citecolor=blue,
    filecolor=magenta,      
    urlcolor=cyan,
}

\begin{document}

\title{Imaging tunable quantum Hall broken-symmetry orders in graphene}

\author{Alexis Coissard}
\thanks{These authors contributed equally to this work.}
\author{David Wander}
\thanks{These authors contributed equally to this work.}
\author{Hadrien Vignaud}
\author{Adolfo G. Grushin}
\affiliation{Univ. Grenoble Alpes, CNRS, Grenoble INP, Institut N\'{e}el, 38000 Grenoble, France}
\author{C\'ecile Repellin}
\affiliation{Univ. Grenoble-Alpes, CNRS, LPMMC, 38000 Grenoble, France}
\author{Kenji Watanabe}
\affiliation{Research Center for Functional Materials, National Institute for Materials Science, 1-1 Namiki, Tsukuba 305-0044, Japan}
\author{Takashi Taniguchi}
\affiliation{International Center for Materials Nanoarchitectonics, National Institute for Materials Science,  1-1 Namiki, Tsukuba 305-0044, Japan}
\author{Fr\'{e}d\'{e}ric Gay}
\author{\linebreak Clemens B. Winkelmann}
\author{Herv\'e Courtois}
\author{Hermann Sellier}
\author{Benjamin Sac\'{e}p\'{e}}
\email{Corresponding author : benjamin.sacepe@neel.cnrs.fr}
\affiliation{Univ. Grenoble Alpes, CNRS, Grenoble INP, Institut N\'{e}el, 38000 Grenoble, France}

\begin{abstract}
\textbf{
When electrons populate a flat band their kinetic energy becomes negligible, forcing them to organize in exotic many-body states to minimize their Coulomb energy \cite{Cao2018a,Cao2018b,Wong2020,Zondiner2020,Saito2021}. The zeroth Landau level of graphene under magnetic field is a particularly interesting strongly interacting flat band because inter-electron interactions are predicted to induce a rich variety of broken-symmetry states with distinct topological and lattice-scale orders \cite{Nomura06,Alicea06,Herbut07a,Jung09,Kharitonov2012,Young2012}. Evidence for these stems mostly from indirect transport experiments that suggest that broken-symmetry states are tunable by boosting the Zeeman energy \cite{Jarillo2014} or by dielectric screening of the Coulomb interaction \cite{Veyrat2020}. However, confirming the existence of these ground states requires a direct visualization of their lattice-scale orders \cite{Li2019}. Here, we image three distinct broken-symmetry phases in graphene using scanning tunneling spectroscopy. We explore the phase diagram by tuning the screening of the Coulomb interaction by a low or high dielectric constant environment, and with a magnetic field. In the unscreened case, we unveil a Kekul\'e bond order, consistent with observations of an insulating state undergoing a magnetic-field driven Kosterlitz-Thouless transition \cite{Checkelsky2008,Checkelsky2009}. Under dielectric screening, a sublattice-unpolarized ground state \cite{Veyrat2020} emerges at low magnetic fields, and transits to a charge-density-wave order with partial sublattice polarization at higher magnetic fields. The Kekul\'e and charge-density-wave orders furthermore coexist with additional, secondary lattice-scale orders that enrich the phase diagram beyond current theory predictions \cite{Nomura06,Alicea06,Herbut07a,Jung09,Kharitonov2012}. This screening-induced tunability of broken-symmetry orders may prove valuable to uncover correlated phases of matter in other quantum materials.
}
\end{abstract}

\maketitle

Narrow electronic energy bands are exceptional playgrounds to explore many-body quantum phases of matter. The vanishingly small kinetic energy in these narrow bands leaves electrons subjected to interaction effects alone, resulting in the emergence of a wealth of correlated, topological and broken-symmetry phases \cite{Cao2018b,Cao2018a,Wong2020,Zondiner2020,Saito2021}. 
Nearly perfect flat bands naturally develop in two-dimensional electron systems under a perpendicular magnetic field, $B$, as macroscopically-degenerate Landau levels. 
There, the main consequence of Coulomb interaction is to generate incompressible --gapped-- phases at half integer filling of Landau levels, by favoring a spin-polarized ground state, a phenomenon called quantum Hall ferromagnetism \cite{Ezawa2013}.

\begin{figure*}[ht]
\includegraphics[width=1\linewidth]{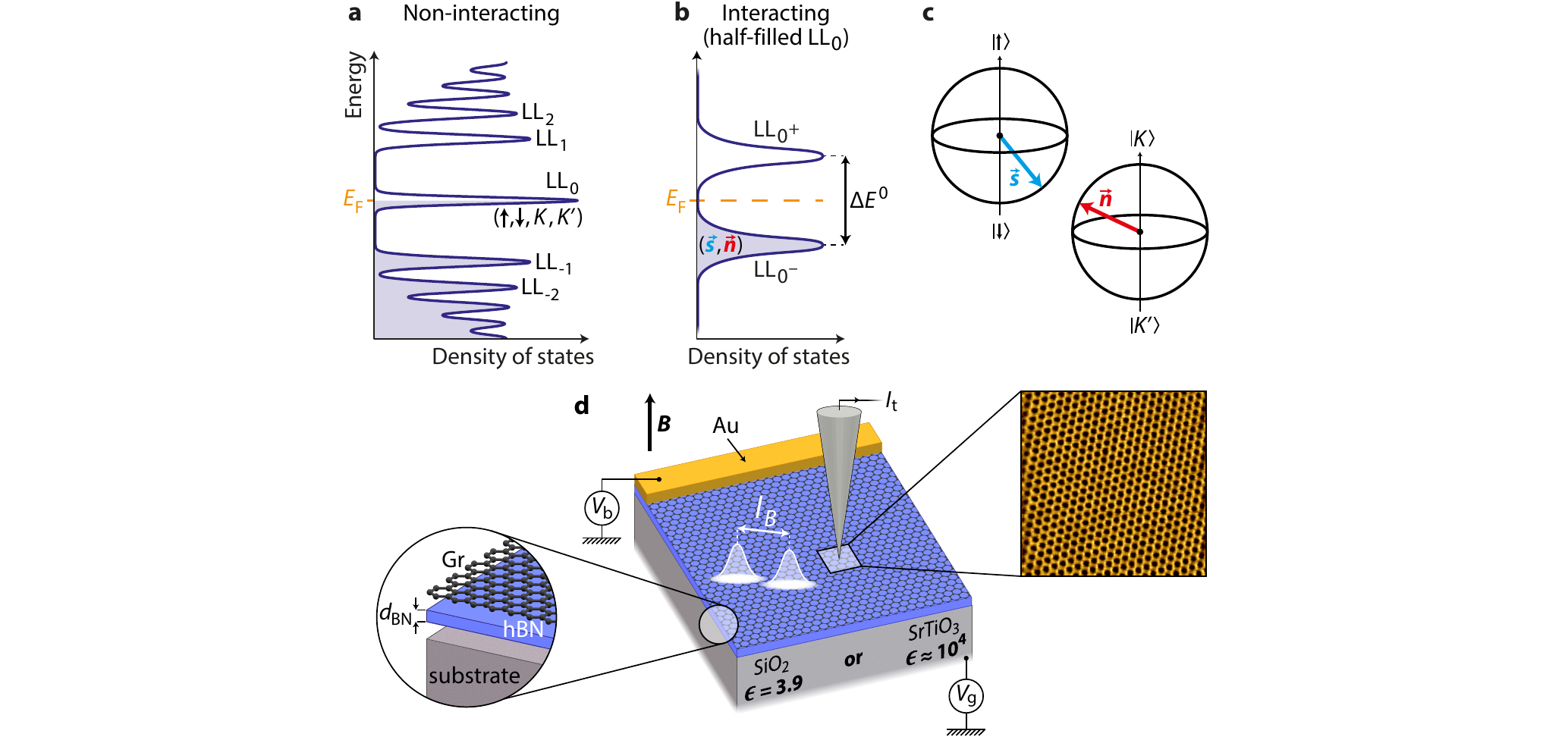}
\centering
\caption{\textbf{Landau level tunneling spectroscopy in graphene.} \textbf{a}, Non-interacting density of states of graphene under perpendicular magnetic field. Each Landau level (LL$_N$, $N$ is the Landau level index) is spin $(\uparrow,\downarrow)$ and valley $(K,K')$ degenerate and emerges as a peak separated from others by cyclotron gaps. \textbf{b}, Due to exchange interaction, the half-filled LL$_0$ at charge neutrality splits into two peaks LL$_{0^\pm}$, thus opening an interaction-induced gap $\Delta E^{0}$. The resulting broken-symmetry state features a SU(4)-polarization with a valley-polarization $\mathbf{n}$ and a spin-polarization $\mathbf{s}$ evolving in the $\text{SU(4)}$ Bloch sphere constructed on the spin and valley subspaces, whose individual Bloch spheres are schematized in \textbf{c}. \textbf{d}, Schematics of the graphene samples. The two different substrates used are either SiO$_2$ (unscreened configuration) or the high-$k$ dielectric SrTiO$_3$ (screened configuration). In both cases graphene is biased with the voltage $V_\text{b}$ through an ohmic contact (in yellow) and its charge carrier density is tuned by the voltage $V_\text{g}$ applied to a back-gate electrode. The tunneling current $I_\text{t}$ is measured from the metallic tip. The tunneling spectroscopy is performed under perpendicular magnetic field $B$. The white disks on the sample illustrate the gaussian electronic wavefunction of LL$_0$ that extends on the scale of the magnetic length $l_B$. The left inset shows the graphene/hBN/substrate heterostructure, where $d_\text{BN}$ is the hBN thickness. The right inset shows a $5\times 5\:\text{nm}^2$ STM image of the graphene honeycomb lattice measured on sample STO07 at 4 K and 0 T.}
\label{Fig1}
\end{figure*}

In graphene, the additional valley degeneracy enriches the quantum Hall ferromagnetism with broken-symmetry states at every quarter filling \cite{Nomura06,Young2012}. A central challenge is to unveil the nature of the ground state of the gapped zeroth Landau level (zLL) at charge neutrality. Theory predicts a rich phase diagram of broken-symmetry states with different topological properties \cite{Alicea06,Herbut07a,Herbut07b,Jung09,Kharitonov2012}. 
While all are SU(4) ferromagnets, their exact spin and valley polarization (Fig.~\ref{Fig1}c) depends on a delicate balance between Zeeman energy and valley-anisotropy terms emerging from the lattice-scale interactions. 
Furthermore, the zLL wavefunctions feature a simple structure in which each valley degree of freedom is locked to one of the graphene's sublattices. This property isolates four possible broken-symmetry states with distinct sublattice or spin orders: a valley-polarized charge-density wave (CDW), a valley-polarized Kekul\'e bond (KB) order, a canted anti-ferromagnet (CAF) and a spin-polarized ferromagnet (F) (see Figs.~\ref{Fig3}a-c). Among them, the F order is a quantum Hall topological insulator harboring conducting helical edge states, while the rest are insulators with gapped edge states \cite{Kharitonov16}.

Experimentally, main insights to distinguish different broken-symmetry states come from transport measurements. High-mobility graphene  typically shows an unequivocal insulating behavior at charge neutrality upon increasing perpendicular magnetic field \cite{Kim2006,Abanin2007,Checkelsky2008,Checkelsky2009,Young2012}. 
On the other hand, a strong in-plane magnetic field that boosts the Zeeman effect and the ensuing spin-polarization can induce a transition to the helical phase with F order \cite{Jarillo2014}. Another alternative strategy utilized a high dielectric constant substrate to screen the long-range part of the Coulomb interaction \cite{Veyrat2020}. This enables the helical phase to emerge at moderate perpendicular magnetic fields, which eventually transits to a weak insulator upon increasing the magnetic field. Besides, a KB order has been observed recently in graphene on graphite, which however cannot be assessed by transport experiments due to the conductive graphite layer \cite{Li2019}. These observations suggest that a broad part of the phase diagram can be explored.

Here, we unambiguously identify three broken-symmetry states in the zLL of graphene by directly visualizing their lattice-scale order with scanning tunneling microscopy (STM) and spectroscopy \cite{Andrei2012}. To access the different broken-symmetry states, we employed two different dielectric materials as substrate, both equipped with a back-gate electrode: the standard silicon oxide (SiO$_2$) and the quantum paraelectric strontium titanate oxide (SrTiO$_3$) with a remarkably high static dielectric constant $\epsilon_\text{STO} \sim 10^4$ at low temperatures (see SI). We fabricated samples consisting of monolayer graphene resting atop a thin hexagonal boron nitride (hBN) flake, deposited on the chosen substrate. To enable screening of the long-range Coulomb interaction \cite{Veyrat2020}, we selected hBN flakes (see SI) with thickness less than or of the order of the magnetic length $l_B = \sqrt{\hbar/eB}$ ($\hbar$ is the reduced Planck constant and $e$ is the electron charge) at low magnetic field, that is, the inter-electron distance in the zLL. Figure~\ref{Fig1}d displays a schematic of the sample structure, where a metallic contact on the graphene serves to apply a voltage bias, $V_\text{b}$. All measurements were performed at 4.2 K.

\begin{figure*}[ht]
\includegraphics[width=1\linewidth]{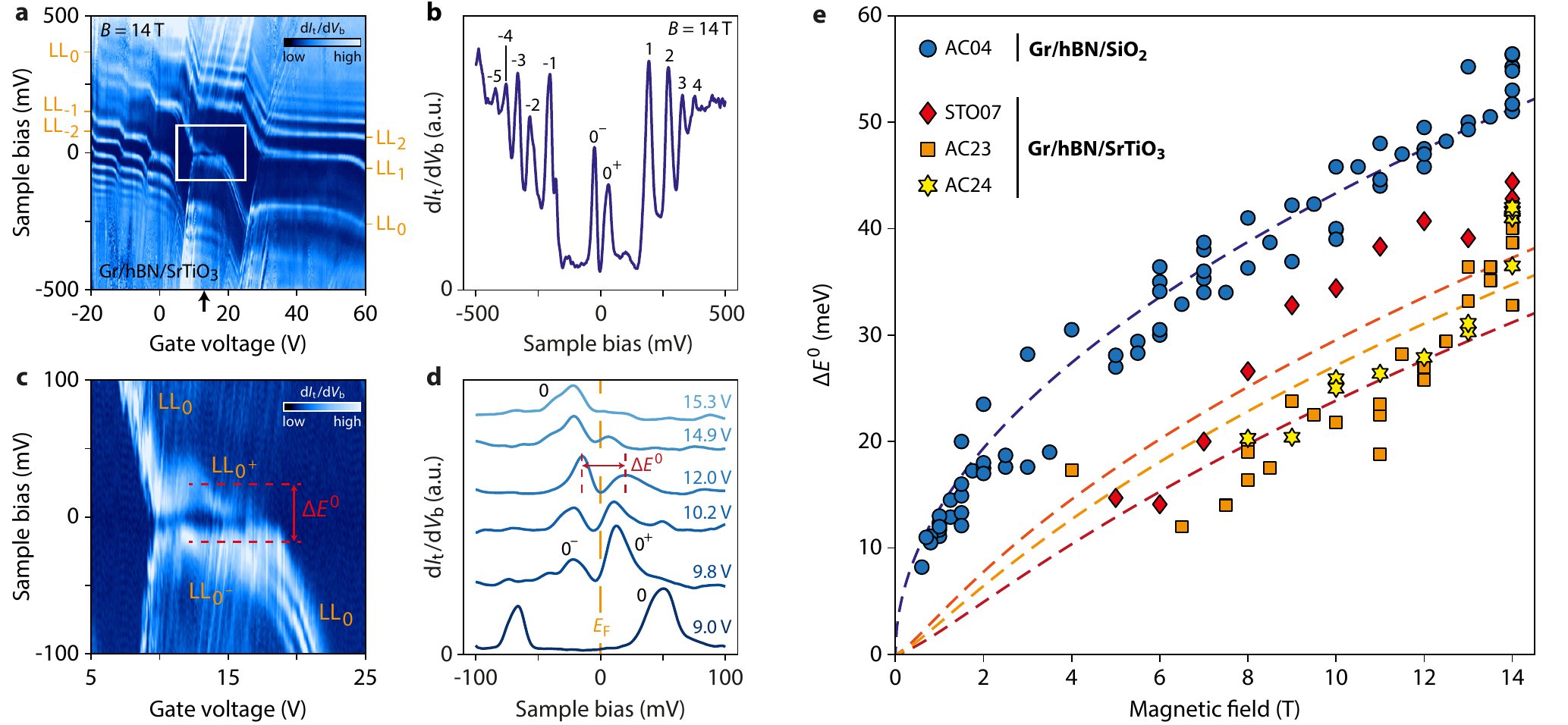}
\centering
\caption{\textbf{Quantum Hall ferromagnetic gap at charge neutrality.} \textbf{a}, Local tunneling conductance gate map measured on graphene sample STO07 (SrTiO$_3$ substrate) at $B=14\:$T. The staircase pattern shows the successive pinning of the Fermi level $E_\text{F}$ inside Landau levels. \textbf{b}, Tunneling spectrum measured on the same sample at $B=14\:$T and $V_\text{g}=13\:$V, which corresponds to charge neutrality as indicated by the black arrow in \textbf{a}. With $E_\text{F}$ at half filling, the zLL splits into two peaks LL$_{0^\pm}$ that define the gap $\Delta E^{0}$ of the broken-symmetry state. \textbf{c}, Zoom of the white rectangle in \textbf{a} showing the splitting of the LL$_0$. \textbf{d}, Spectra extracted from \textbf{c} at the back-gate voltages indicated on the right of each curve. \textbf{e}, Evolution of $\Delta E^{0}$ as a function of magnetic field for the four studied samples. For sample AC04 (SiO$_2$ substrate) the gap is fitted by the Coulomb energy $1/2\sqrt{\pi/2}\:\mathcal{E}_\text{C}\propto\sqrt{B}$ shown as a dashed blue line. For the three samples with SrTiO$_3$ substrates, $\Delta E^{0}$ is decreased compared to sample AC04 due to the substrate screening of the Coulomb interaction. The red, orange and yellow dashed lines correspond to the substrate-screened Coulomb energy $1/2\sqrt{\pi/2}\:\tilde{\mathcal{E}}_\text{C}$, computed with the respective hBN thickness of the samples.}
\label{Fig2}
\end{figure*} 

\section*{Coulomb interaction screening}

The Coulomb interaction strength can be readily assessed by tunneling spectroscopy of the exchange gap that opens at half filling of the zLL \cite{Li2019}  (see Figs.~\ref{Fig1}a and b). Figure~\ref{Fig2}b displays a representative local tunneling conductance spectrum, $\text{d}I_\text{t}/\text{d}V_\text{b}$ versus $V_\text{b}$, measured on sample STO07 under a perpendicular magnetic field of 14~T. In this measurement, the Fermi level is adjusted at charge neutrality, that is, at half filling of the zLL, by applying a back-gate voltage $V_\text{g}=13$~V. While all Landau levels with index $|N|\geq1$ appear in the tunneling conductance as sharp peaks separated by the cyclotron gap scaling as $\sqrt{|N|B}$ (see SI), the zLL splits into two peaks revealing the Coulomb gap of the $\nu=0$ broken-symmetry state, $\Delta E^{0}$, akin to earlier experiments in GaAs \cite{Dial2007}. For accurate measurements of the gap, which reaches its maximum value at half-filling \cite{Ezawa2013}, we measured the back-gate dependence of the tunneling conductance as shown in Fig.~\ref{Fig2}a. The Landau level peaks form a staircase pattern, indicating the successive pinning of the Fermi level within each highly-degenerate Landau level \cite{Luican2011,Chae2012}. The absence of Coulomb diamond features seen in previous studies indicates that our measurements are not affected by a tip-induced quantum dot \cite{Jung11}. Analysis of individual spectra around charge neutrality enables us to evaluate $\Delta E^{0}$, defined as the maximum separation between split peaks (red arrow in Figs.~\ref{Fig2}c and d). Note that tip-induced gating yields a negligible variation of filling factors of the order of 0.1 at the bias voltage of the split peaks (see SI).

Inspecting a systematic set of conductance maps for different magnetic field values, measured at various locations on the graphene surface, and on four different samples including SiO$_2$ and SrTiO$_3$ substrates (see SI), provides a robust determination of the $B$-dependence of the energy gap. Figure~\ref{Fig2}e displays the resulting values of $\Delta E^{0}$ as a function of magnetic field. 

We first focus on the unscreened case of sample AC04 (SiO$_2$ substrate) with the blue data points. A clear $\sqrt{B}$ dependence highlighted by the blue dashed line is observed starting at fields as low as $0.6$~T and up to $14$~T. This dependence reflects the growth of the Coulomb energy with $B$ that scales as $\mathcal{E}_\text{C}=e^2 / 4\pi \epsilon_0 \epsilon_\text{r} l_B\propto\sqrt{B}$, where $\epsilon_0$ and $\epsilon_\text{r}$ are the vacuum permittivity and the relative permittivity surrounding the graphene.  As the top graphene surface is exposed to vacuum, $\epsilon_\text{r} = (\epsilon_\text{BN}+1)/2 \simeq 2.3$, where $\epsilon_\text{BN}\simeq 3.6$ is the hBN relative permittivity. Theoretically, $\Delta E^{0}$ is expected to be $1/2\sqrt{\pi/2}\:\mathcal{E}_\text{C}$ \cite{Ezawa2013}. 
We plot this quantity in Fig.~\ref{Fig2}e (blue dashed curve) by adjusting $\epsilon_\text{r}$ to $2.6$, which is consistent with the expected value for the relative permittivity. Such a quantitative agreement demonstrates the significance of our spectroscopy to assess the interaction-induced gap.

\begin{figure*}[!ht]
\includegraphics[width=1\linewidth]{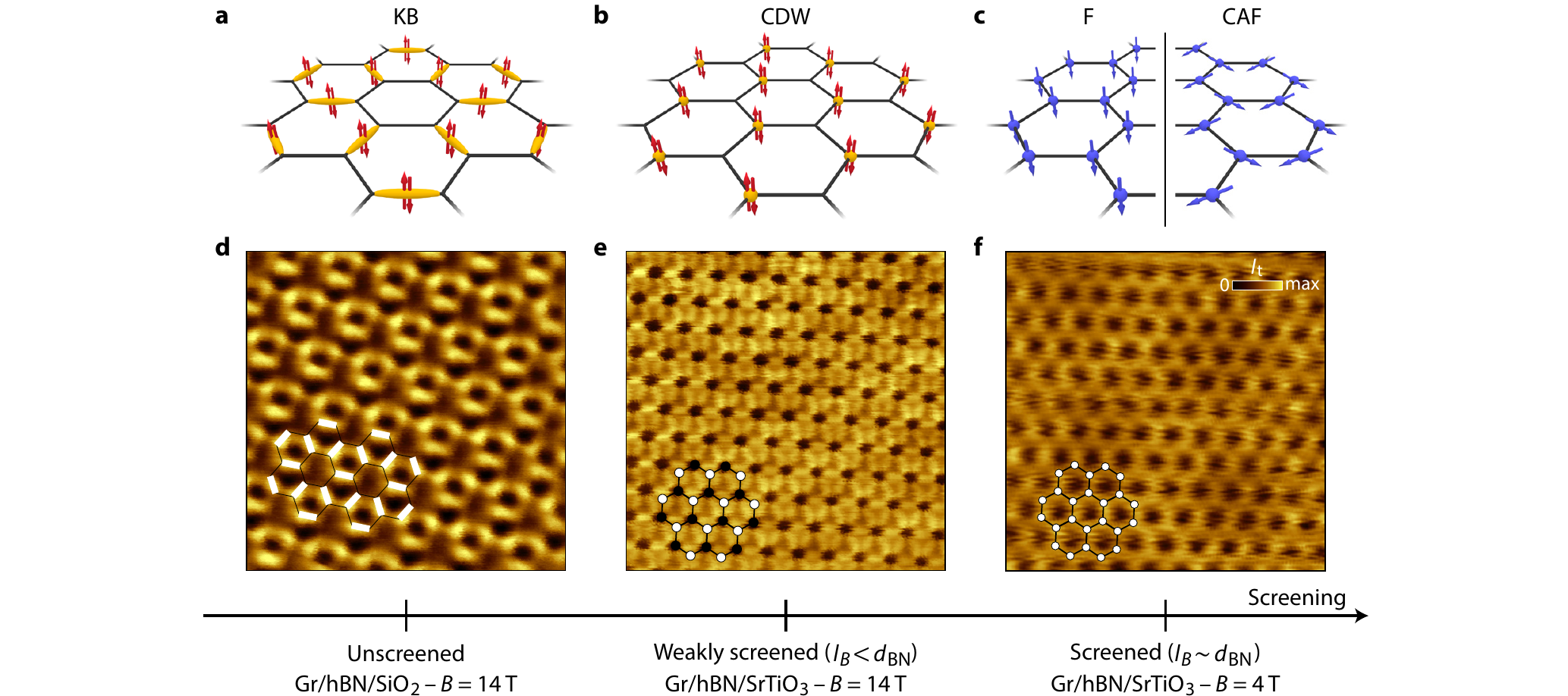}
\centering
\caption{\textbf{Tunable broken-symmetry states of charge-neutral graphene.}  \textbf{a-c}, Lattice-scale order drawings of the four possible broken-symmetry states in charge-neutral graphene under perpendicular magnetic field. \textbf{a} shows the Kekul\'e bond (KB) order, \textbf{b}, the charge-density wave (CDW) with sublattice polarization and \textbf{c}, the spin-polarized ferromagnetic (F) and canted anti-ferromagnetic (CAF) ground states. \textbf{d-f}, $2.6\times 2.6\:$nm$^2$ STM images taken in constant height mode. \textbf{d}, STM image on sample AC04 (unscreened graphene) at $B=14\:$T and $V_\text{b}=25\:$mV, which unveils a Kekul\'e bond order. \textbf{e}, At $B=14\:$T (weakly-screened graphene, $l_B=7\:\text{nm}<d_\text{BN}=12\:\text{nm}$) and $V_\text{b}=40\:$mV, the sample AC23 exhibits a sublattice charge-density wave, whereas \textbf{f}, at $B=4\:$T (screened graphene, $l_B=13\:\text{nm}\sim d_\text{BN}=12\:\text{nm}$) and $V_\text{b}=20\:$mV, we find a valley-unpolarized phase consistent with the spin-polarized helical phase. For each image we superposed a corresponding lattice drawing emphasizing the bond order, the sublattice polarization and the honeycomb lattice, respectively. The black horizontal arrow indicates the strength of substrate screening.}
\label{Fig3}
\end{figure*} 

Remarkably, turning to the screened case with the SrTiO$_3$ substrate yields gap values conspicuously smaller than those obtained on the sample on SiO$_2$ (see red, orange and yellow data points in Fig.~\ref{Fig2}e). This demonstrates a clear screening of the Coulomb interaction by the high-dielectric constant of the substrate. Electrostatic considerations that account for the thin hBN bottom layer lead to a substrate-screened Coulomb energy scale $\tilde{\mathcal{E}}_{\text{C}}=\mathcal{E}_{\text{C}} \times S(B)$ that is mitigated by a screening factor $S(B)\approx 1-\frac{\epsilon_{\text{STO}}-\epsilon_{\text{r}}}{\epsilon_{\text{STO}}+\epsilon_{\text{r}}}\frac{l_B}{\sqrt{l_B^2+4d_{\text{BN}}^2}}$, where $d_{\text{BN}}$ is the hBN thickness \cite{Veyrat2020}. Consequently, electrons in graphene are subjected to an unusual $B$-dependent screening that depends on the ratio $l_{B}/d_{\text{BN}}$ and is most efficient at low magnetic fields. 
In Fig.~\ref{Fig2}e the red, orange and yellow dashed curves show $\tilde{\mathcal{E}}_{\text{C}}$ calculated with the hBN thickness of the respective samples. Although the use of $\tilde{\mathcal{E}}_{\text{C}}$ is strictly valid only for hBN-encapsulated graphene, we obtain a decent agreement with our data, despite some scattering for sample STO07.

\begin{figure*}[!ht]
\includegraphics[width=1\linewidth]{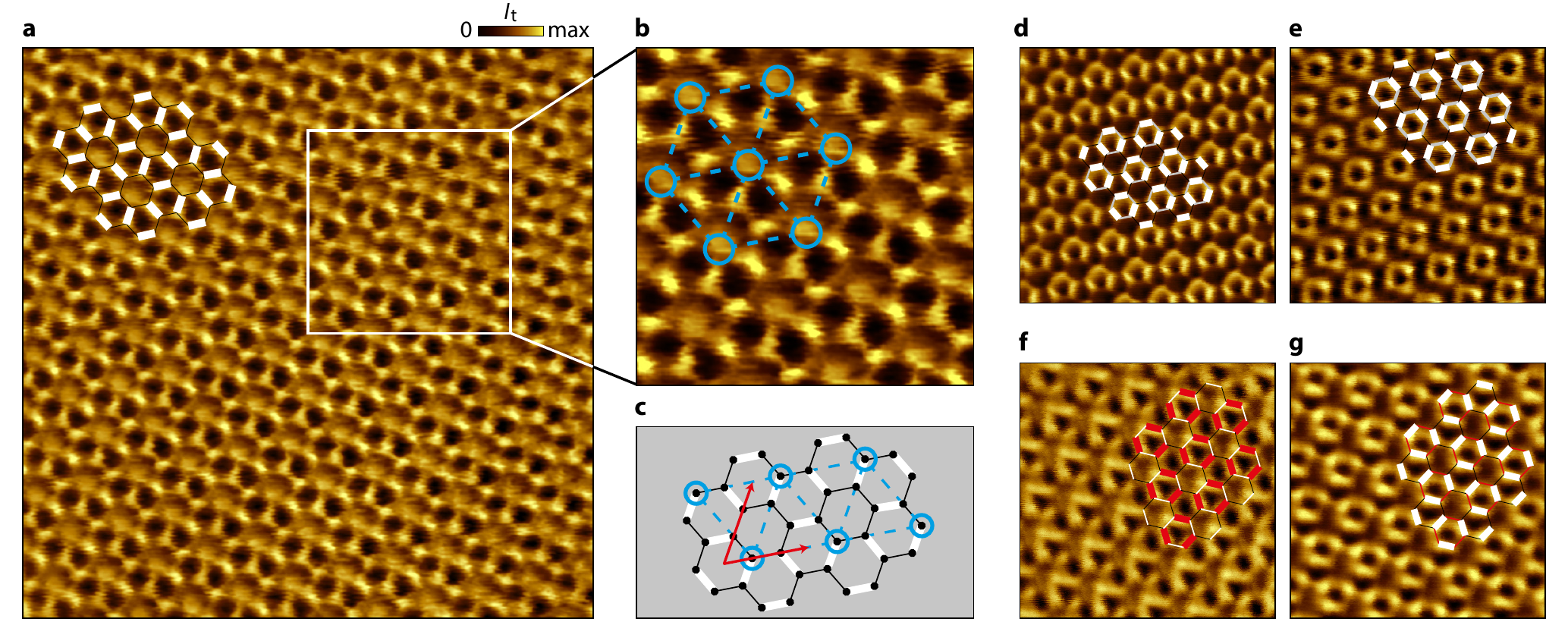}
\centering
\caption{\textbf{Kekul\'e bond order in unscreened charge-neutral graphene.} \textbf{a}, $5\times 5\:$nm$^2$ STM image on sample AC04, at $B=14\:$T and $V_\text{b}=25\:$mV. The KB lattice, where both electrons of the broken-symmetry state are localized on one C-C bond out of three, is shown in overlay with the stronger bonds highlighted in thick white. This structure is detailed in \textbf{c}, where the basis vectors of the KB order lattice are indicated in red. A zoom of \textbf{a} is displayed in \textbf{b}, where we observe a secondary charge-density-wave order featuring a triangular lattice (drawn in blue) with a parameter matching the one of the Kekul\'e lattice ($\sqrt{3}$ times the graphene lattice parameter). \textbf{d, e}, $3\times 3\:$nm$^2$ images showing asymmetric Kekul\'e patterns. \textbf{d}, At $B=14\:$T and $V_\text{b}=2\:$mV the three Kekul\'e strong bonds are partially merged, while in \textbf{e}, at $B=3\:$T and $V_\text{b}=30\:$mV, they are completely merged and form a circle-like pattern. \textbf{f, g}, $3\times 3\:$nm$^2$ images at $B=14\:$T, at $V_\text{b}=25\:$meV acquired at the same position a few minutes apart, showing the transition between two degenerate Kekul\'e configurations from the red Kekul\'e lattice in \textbf{f}, to the white one in \textbf{g}. Both lattices in overlay are at the same position.}
\label{Fig4}
\end{figure*} 

\section*{Tunable lattice-scale orders}

We now turn to the central result of this work, benchmarking the lattice-scale orders of the charge-neutral broken-symmetry state, upon tuning the screening of the Coulomb interaction.
Figure~\ref{Fig3} shows three STM images taken at the energy of a split zLL peak, on the SiO$_2$ sample AC04 at $B=14$~T  (Fig.~\ref{Fig3}d), and on the SrTiO$_3$ sample AC23 at $B=14$~T (Fig.~\ref{Fig3}e) and $4$~T (Fig.~\ref{Fig3}f). These panels thus cover three regimes for Coulomb interaction that we qualify as unscreened, moderately screened and screened, respectively. For the unscreened case in Fig.~\ref{Fig3}d, we observe a Kekul\'e distortion bond-order pattern of the electronic wavefunction, consistent with an independent recent study \cite{Liu2021}, indicating that spin-singlet pairs of electrons are localized on one bond out of three per carbon atom of the graphene honeycomb lattice. This order is stable down to $B=3$~T (Fig.~\ref{Fig4}e).  With the SrTiO$_3$ substrates at high magnetic field, that is, under moderate screening, another lattice-scale order develops with a stark valley-polarization: the CDW ground-state with the spin-singlet pairs now mostly localized on a single sublattice (Fig.~\ref{Fig3}e). This CDW order is found to be independent of the presence of a Moir\'e superlattice formed with the hBN layer (see SI). Finally, at low magnetic field (4~T in Fig.~\ref{Fig3}f), this CDW order disappears, revealing a valley-unpolarized graphene honeycomb lattice, which points to a spin order, that is, the F or CAF orders.

\begin{figure*}[!ht]
\includegraphics[width=1\linewidth]{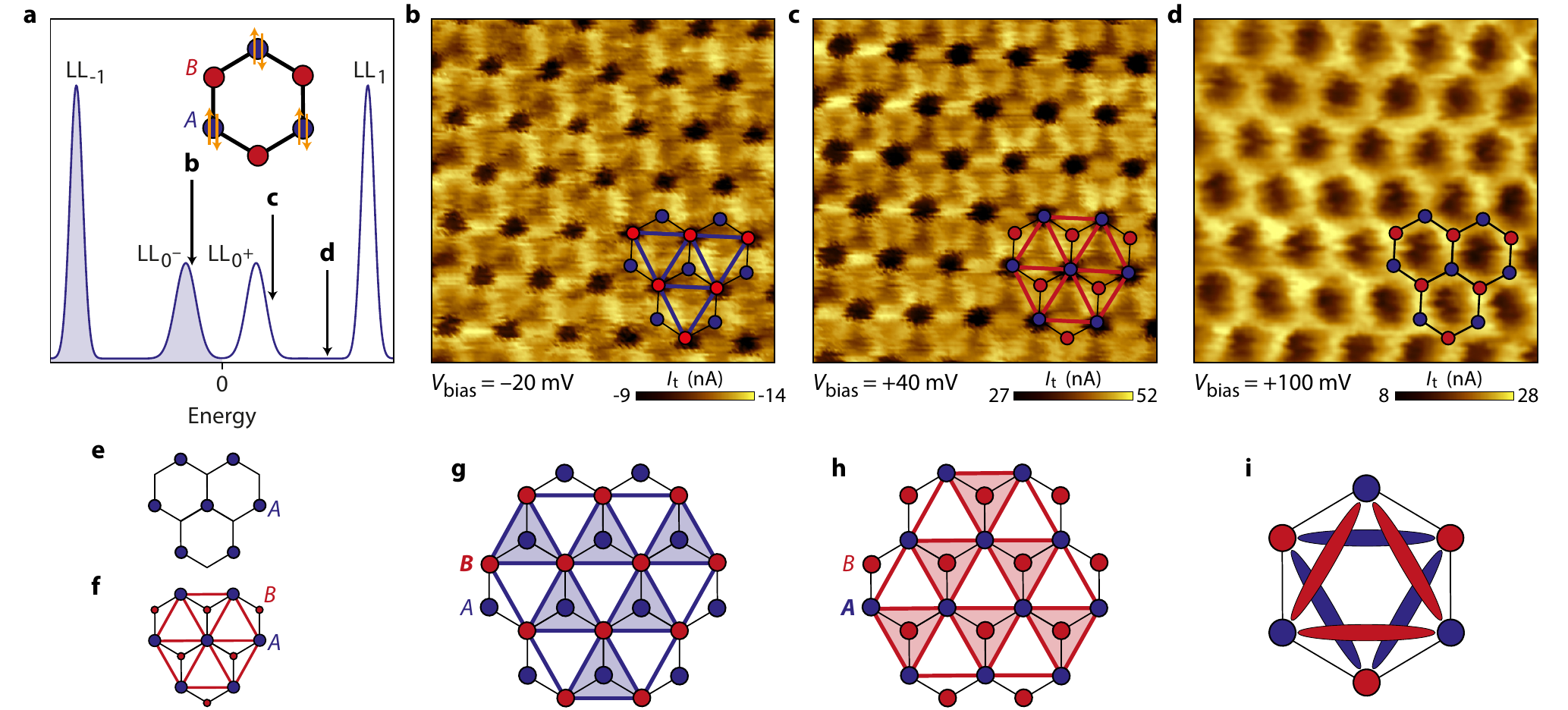}
\centering
\vspace{-1.6em}
\caption{\textbf{Charge-density-wave order in moderately screened charge-neutral graphene.} \textbf{a}, Simulated Landau spectrum indicating the bias voltage positions of the subsequent $1.5\times 1.5\:$nm$^2$ constant-height STM images, all acquired at $B = 14$~T and at the same location on sample AC23. Inset:  Convention for the CDW lattice, the blue sublattice $A$ is doubly occupied whereas the red sublattice $B$ is empty. \textbf{b}, Image at $V_\text{b}=-20\:$mV with the CDW lattice in overlay: the sublattice $B$ (empty) appears as dark spots at negative bias and intra-sublattice bonds are visible as bright lines between them. \textbf{c}, Similarly, at the positive bias of $V_\text{b}=+40\:$mV, the sublattice $A$ (doubly occupied) appears as dark spots and intra-sublattice bonds are visible as bright lines between them. Note that the colorscale in \textbf{b} is inverted with respect to that in \textbf{c}, since the tunneling current is negative due to the negative bias.  
 \textbf{d}, At $V_\text{b}=100\:$mV, the CDW is no longer visible and the honeycomb lattice appears instead. \textbf{e}, CDW with full sublattice polarization, as predicted in Ref. \cite{Kharitonov2012}, compared to a CDW with partial sublattice polarization in \textbf{f}. The symmetry-allowed triangular bond order is suppressed in \textbf{e}, and coexists with the CDW in \textbf{f}, where a triangular lattice emerges due to a symmetry allowed sublattice hopping-asymmetry. In \textbf{f}, the difference in size for the $A$ and $B$ atoms represents the partial sublattice polarization.
 \textbf{g, h}, Sketch showing the CDW sublattice inversion between negative bias in \textbf{g}, and positive bias in \textbf{h}. \textbf{i}, Structure of the intra-sublattice bonds.}
\label{Fig5}
\end{figure*} 

\section*{Kekul\'e bond-order}

Going further, we show that additional fine structures emerge at the lattice scale, enriching the predicted phase diagram. 
The KB order features an unexpected, faint charge-density wave that has the periodicity of the Kekul\'e unit cell. 
This coexisting order is readily seen in Fig.~\ref{Fig4}a : a nicely formed Kekul\'e pattern exhibits an enhanced local wavefunction amplitude inside one hexagon of the honeycomb that repeats periodically on the Kekul\'e triangular lattice, as indicated by the blue circles and dashed lines in Fig.~\ref{Fig4}b. Figure~\ref{Fig4}c provides a representation of the latter superimposed on the honeycomb and Kekul\'e lattices.
This new charge-density wave that we label as K-CDW is different from the CDW broken-symmetry state since it displays a triangular lattice with a parameter $\sqrt{3}$ times larger than the graphene lattice parameter. The tripled unit cell of the K-CDW is reminiscent of CDW phases observed to compete with the Kekul\'e order in extended Hubbard-models \cite{Motruk2015,Capponi2015} at $B=0$, but have not been reported or predicted at finite $B$. 
We also observed other situations in which this K-CDW induces a pronounced asymmetry of the Kekul\'e pattern, with a more or less merging of the Kekul\'e strong bonds (see SI). This is illustrated by the evolution from the mostly symmetric Kekul\'e lattice shown in Fig.~\ref{Fig4}a to the asymmetric Kekul\'e lattices shown in Figs.~\ref{Fig4}d and e. If on some images the strong bonds are still visible (see Fig.~\ref{Fig4}d), they can also merge with each other, forming a circle-like pattern seen in Fig.~\ref{Fig4}e.

Interestingly, both the KB order and K-CDW vary with time. For instance, Figs.~\ref{Fig4}f and g show a spontaneous transition from one of the three possible degenerate Kekul\'e lattices to another, while imaging continuously the very same location. Similar changes for the K-CDW are shown in SI.

\section*{Charge-density-wave order}

In screened graphene, likewise, close inspection of the CDW broken-symmetry state reveals striking fine structures. Here, we assume that the electron doublets of the CDW are localized on the sublattice $A$ (in blue), while the sublattice $B$ is empty (in red) as in the inset of Fig.~\ref{Fig5}a. We start by deciphering Figs.~\ref{Fig5}b and c, taken at the exact same location, and comparing the occupied and empty orbitals of the same atoms. In both images the CDW appears as dark spots featuring a triangular symmetry and corresponding to the atoms of a single sublattice (the sublattice $B$ in Fig.~\ref{Fig5}b and the sublattice $A$ in Fig.~\ref{Fig5}c), whereas atoms of the other sublattice are not visible. Extracting electrons from the occupied states into the STM tip at negative sample bias leads to a low tunneling current on the empty atoms of the sublattice $B$ (dark spots in Fig.~\ref{Fig5}b). Similarly, injecting electrons from the STM tip into the empty states at positive sample bias leads to a low current on the already doubly occupied atoms of the sublattice $A$ (dark spots in Fig.~\ref{Fig5}c).

In addition, inspecting Fig.~\ref{Fig5}c, we see that atoms of sublattice $B$ (red spots in the inset) are located in the middle of a triangular lattice formed by bright lines joining the doubly-occupied atoms of the sublattice $A$, which correspond to high density of empty states. 
One can also notice in Fig.~\ref{Fig5}c that the triangular cells that contain atoms $B$ are brighter than the other empty cells, which is consistent with the enhancement of the local density of empty states by the $B$ atoms. 
We illustrate this unusual sublattice inversion with the lattice drawings in Figs.~\ref{Fig5}g and h, where the density of occupied and empty states are color coded blue and red, respectively. 
Figure~\ref{Fig5}i summarizes this remarkable inverted pattern by superposing the empty (red) and occupied (blue) states on the same carbon hexagon.

This additional triangular order that accompanies the CDW in Fig.~\ref{Fig5} was not expected by theory \cite{Kharitonov2012}.
Such a sublattice bond asymmetry, $t_{2,A} \neq t_{2,B}$ ($t_{2,A}$ and $t_{2,B}$ are the second-nearest-neighbor bond expectation values of sublattices $A$ and $B$, respectively), is generically permitted in the CDW phase as it gaps out the zLL with the same matrix structure as a sublattice charge imbalance \cite{Alba2011} (see SI). It becomes visible when the sublattice degree of freedom is partially polarized (see Figs.~\ref{Fig5}e and f).  We conjecture that this partial sublattice polarization could originate from Landau level mixing, since the sublattice index is decoupled from the valley index in higher Landau levels \cite{PetersonNayakPRL2014, Feshami_2016,Das2021}.

\section*{Implications on transport experiments}

The observation of these three ground states has profound implications for the understanding of transport properties at charge neutrality. For the unscreened case, corresponding to virtually all transport experiments, the KB order contradicts the transition scenario from the conjectured CAF phase to the helical F phase tuned by the Zeeman field \cite{Kharitonov2012,Jarillo2014}, as well as with recent magnon transmission experiments \cite{Takei2016,Wei2018,Stepanov2018,Assouline2021} that imply magnetism. Nonetheless, a recent prediction \cite{Das2021} suggests that both KB order and CAF phases could co-exist, thus accounting for the experimental dissonance. Our observation of a sub-dominant K-CDW order adds yet a new flavor to the phase diagram, which was not anticipated thus far and deserves further theoretical attention. Similarly, the nature of edge excitations may be more complex than initially thought \cite{Knothe15}. A definitive conclusion on the existence of an underlying magnetism in this ground state and its possible spin texture at the lattice scale (Fig.~\ref{Fig3}c) would require further spin-filtered scanning tunneling experiments. 

Furthermore, our observation of coexisting orders implies that the nature of bulk excitations in this insulating phase must be revisited \cite{Atteia2021}. The time-varying nature of the KB order and K-CDW, indicating some depinning mechanisms and the presence of domain walls, may contribute to charge transport, in parallel to skyrmion excitations. A Kosterlitz-Thouless phase transition driven by topological Kekul\'{e} vortex zero-modes \cite{Hou2007} has been predicted \cite{Nomura2009} and discussed experimentally \cite{Checkelsky2008,Checkelsky2009}.

In the screened case, we observed the CDW persisting from $B=14\:$T down to $B=7\:$T (see SI) and disappearing at $B=4\:$T in favour of a valley-unpolarized phase at lower magnetic field. This phase transition is consistent with the spin-polarized helical phase observed in transport measurements \cite{Veyrat2020}, which is replaced by a weakly insulating phase at high fields, corresponding to the CDW phase unveiled in this work. The scenario explaining the change of ground state due to substrate screening was accounted for by a modification of renormalization effects of the valley-anisotropy energies by the (screened) long-range Coulomb interaction \cite{Veyrat2020}. Although it is difficult to assess this renormalization experimentally, the substrate screening of the Coulomb interaction evidenced by our spectroscopy of the zLL gap indicates that such a mechanism is likely to be at play. As the screening exhibits an inversely proportional $B$-dependence, our findings, supported by transport experiments, confirm a modification of the valley-anisotropy energies upon varying the magnetic field in this specific substrate-screened configuration.

Ultimately, at much higher fields such that $l_{B}\ll d_{\text{BN}}$, one should expect in graphene samples on SrTiO$_3$ another transition from the CDW to the KB phase. In this situation the substrate screening vanishes and the Coulomb energy scale asymptotically reaches its bare value.

\section*{Data availability}

All data described here are available at Zenodo \cite{Data_zenodo}.

\section*{Acknowledgements}

We thank F. de Juan, H. Fertig, M. Goerbig, G. Murthy and E. Shimshoni for valuable discussions. We thank B. Kousar for careful reading of the manuscript. We thank D. Dufeu, Ph. Gandit, D. Grand, D. Lepoittevin, J.-F. Motte, P. Plaindoux and L. Veyrat for technical assistance in setting up the experimental system. Samples were prepared at the Nanofab facility of the N\'eel Institute. This work has received funding from the European Union's Horizon 2020 research and innovation program ERC grants \textit{QUEST} No. 637815 and \textit{SUPERGRAPH} No. 866365, and the Marie Sklodowska-Curie grant QUESTech No. 766025. A.~G.~G. acknowledges financial support by the ANR under the grant ANR-18-CE30-0001-01~(TOPODRIVE).

\section*{Competing Interests} 

The authors declare that they have no competing financial interests.

\onecolumngrid
\vspace{2em}
\twocolumngrid

\bibliographystyle{naturemag}
\bibliography{MBGS-BIB}

\end{document}


\setcounter{page}{9}

\title{\Large Supplementary Information for \\ Imaging tunable quantum Hall broken-symmetry orders \\ in graphene}

\author{Alexis Coissard$^{1,*}$, David Wander$^{1,*}$, Hadrien Vignaud$^{1}$, Adolfo G. Grushin$^{1}$, C\'ecile Repellin$^{2}$, Kenji Watanabe$^{3}$, Takashi Taniguchi$^{4}$, Fr\'{e}d\'{e}ric Gay$^{1}$, \\Clemens B. Winkelmann$^{1}$, Herv\'e Courtois$^{1}$, Hermann Sellier$^{1}$ \& Benjamin Sac\'{e}p\'{e}$^{1,\,\textrm{\Letter}}$}

~\vfill

\maketitle

\begin{affiliations}
\centering
 \item Univ. Grenoble Alpes, CNRS, Grenoble INP, Institut N\'{e}el, 38000 Grenoble, France
 \item Univ. Grenoble-Alpes, CNRS, LPMMC, 38000 Grenoble, France
 \item Research Center for Functional Materials, National Institute for Materials Science, 1-1 Namiki, Tsukuba 305-0044, Japan
 \item International Center for Materials Nanoarchitectonics, National Institute for Materials Science,  1-1 Namiki, Tsukuba 305-0044, Japan
 \item[*] These authors contributed equally to this work.
 \item[$^\textrm{\Letter}$] email: benjamin.sacepe@neel.cnrs.fr
\end{affiliations}

\vfill
\newpage


\section{Sample fabrication}
	
Graphene/hBN heterostructures were assembled from exfoliated flakes with the van der Waals pick-up technique using a polypropylene carbonate (PPC) polymer \cite{Wang2013}. Stacks were deposited using the methods described in Ref. \cite{Vitto2017} (for sample STO07) or in Ref. \cite{Nadj-Perge2019} (for samples AC04, AC23 and AC24), on either highly doped Si wafers with a $285$~nm thick SiO$_2$ layer, or on 500$\:\mu$m thick SrTiO$_3$ (100) substrates cleaned with hydrofluoric acid buffer solution before deposition of the graphene/hBN heterostructures (a Ti/Au bilayer was deposited later on the other side of the SrTiO$_3$ substrate to enable back-gate effect). The geometrical parameters of the samples are listed in Table~\ref{ExDTab1}. Electron-beam lithography using a PMMA resist was used to pattern a guiding markerfield on the whole $5\times 5\:$mm$^2$ substrate to drive the STM tip toward the device. Cr/Au electrodes contacting the graphene flake were also patterned by electron-beam lithography and metalized by e-gun evaporation. Samples were thermally annealed at $350\,^{\circ}$C in vacuum under an halogen lamp to remove resist residues and clean graphene, before being mounted into the STM where they were heated \textit{in situ} during the cooling to $4.2\;$K.

	
\section{Measurements}
	
Experiments were performed with a home-made hybrid scanning tunneling microscope (STM) and atomic force microscope (AFM) operating at a temperature of $4.2\;$K in magnetic fields up to $14\;$T. The AFM mode is used to guide the tip toward the graphene device. The sensor consists of a hand-cut PtIr tip glued on the free prong of a tuning fork, the other prong being glued on a Macor plate. Once mounted inside the STM, the tip is roughly aligned over the sample at room temperature and then guided toward the graphene in AFM mode at low temperature using the guiding markerfield. Scanning tunneling spectroscopy was performed using a lock-in amplifier technique with a modulation frequency of $263\;$Hz and rms modulation voltage between $1-5\;$mV depending on the spectral range of interest.
Imaging of the zLL lattice-scale orders was carried out in STM constant-height mode. Starting from a tunneling contact at ($V_\text{b}=300\:$mV, $I_\text{t}=1\:$nA) with the Z-regulation on, we switch off the regulation and lower the bias voltage to either energies corresponding to the LL$_{0^\pm}$ peaks, which drastically decreases the tunneling current. We then manually approach the tip toward graphene until the recovery of a tunneling current of a few nA. STM images of the tunneling current measured while scanning at constant tip-sample distance subsequently yield atomically resolved images of the honeycomb lattice or lattice-scale orders.


\section{Samples summary}
\label{secSamples}

\begin{figure}[ht!]
	\centering
	\includegraphics[width=0.7\linewidth]{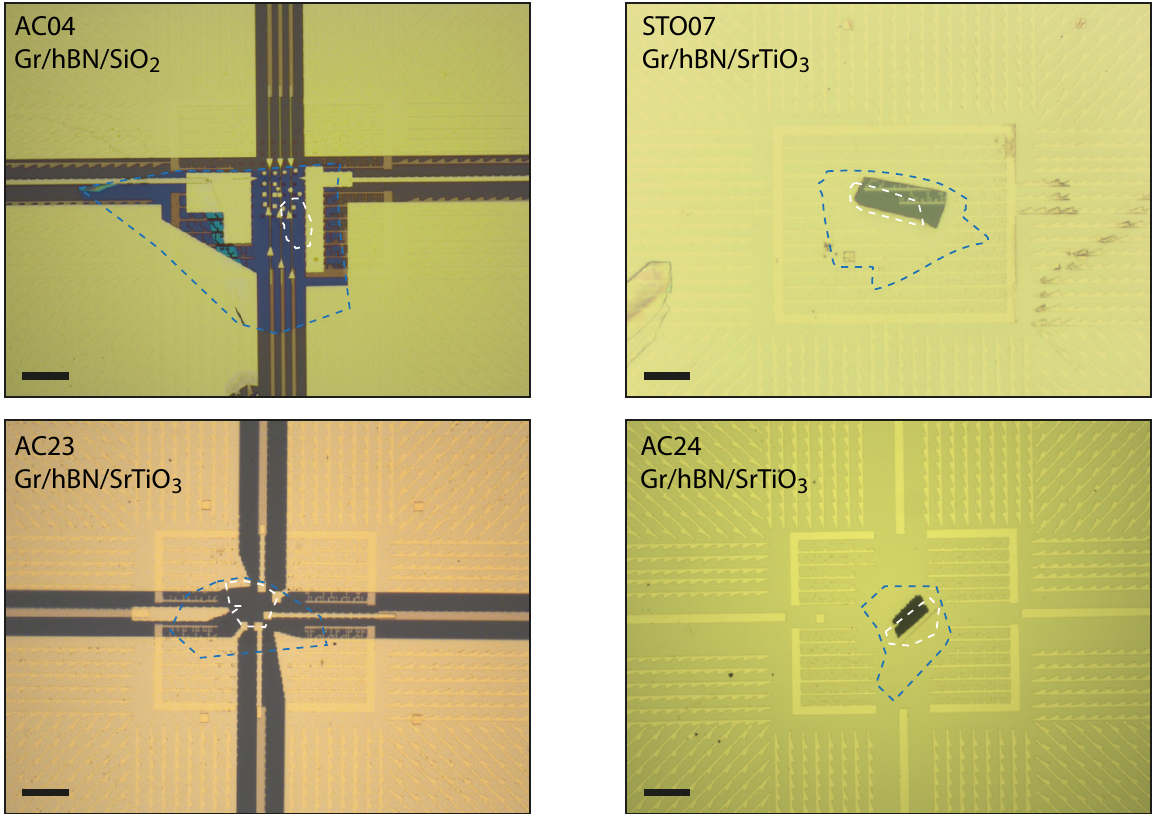}
	\caption{\textbf{Graphene samples.} Optical pictures of the studied samples listed in Table~S1. The dashed blue lines outline the hBN flakes, while the white dashed lines outline the graphene flakes. For every image, the scale bar is $10\:\mu$m.}
	\label{figS1}
\end{figure}

We studied two different types of samples (see Fig.~1d of the main text) : first graphene on hBN/SiO$_2$ substrate, the usual substrate used in transport measurements, and secondly graphene on hBN/SrTiO$_3$ substrate with a thin hBN layer of thickness $\dbn$. Optical images of the samples are shown in Fig. \ref{figS1} and their geometrical parameters are listed in Table \ref{ExDTab1}. Both types of samples are equipped with a back-gate electrode.

~\vfill

\begin{table}[ht!]
	\centering
		\includegraphics[width=0.7\textwidth]{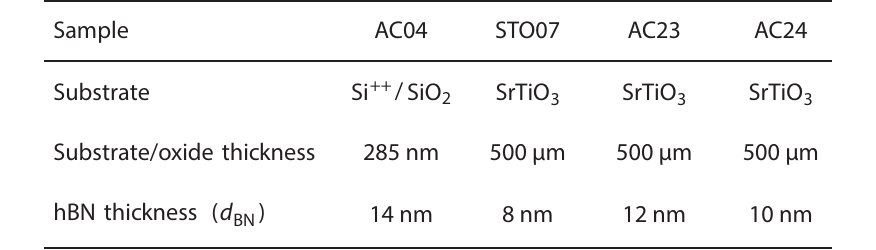}
	\caption{\textbf{Geometrical parameters of the four measured samples.}}
	\label{ExDTab1}
\end{table}

~\vfill

~\vfill


\section{Dielectric constant of S\lowercase{r}T\lowercase{i}O$\mathbf{_3}$ in magnetic field.}

\begin{figure}[!b]
	\centering
		\includegraphics[width=1\textwidth]{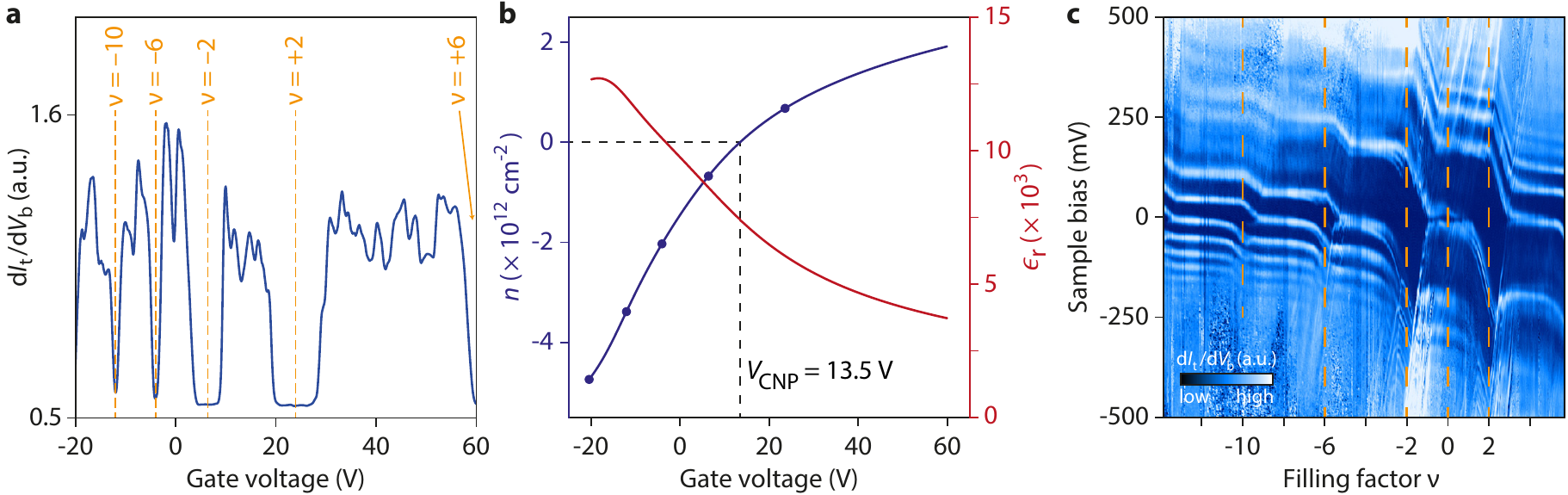}
	\caption{\textbf{Estimation of the dielectric constant of SrTiO$\bm{_3}$ and rescaling of the gate map.} \textbf{a}, Line  cut at $V_\text{b}=0\:$V, averaged on a range of $\pm 20\:$mV, of the d$I_\text{t}$/d$V_\text{b}$ gate map in Fig. 2a. \textbf{b}, Estimation from the filling factors obtained in \textbf{a} of the charge carrier density $n$ (blue dots), its polynomial fit (blue curve), and computed values of $\epsilon_\text{r}\simeq\epsilon_\text{STO}$ (red curve), as a function of gate voltage. The fit yields $V^{\text{CNP}}\simeq 13.5\:$V. \textbf{c}, Rescaling of the gate map of Fig. 2a as a function of $\nu$.}
	\label{ExDFig1}
\end{figure}

We show here how to estimate the SrTiO$_3$ dielectric constant, $\epsilon_\text{STO}$, from tunneling conductance gate maps. $\epsilon_\text{STO}$ is related to the global dielectric constant of the back gate, $\epsilon_\text{r}$, which can be obtained by modeling the back-gate capacitance $C_{\text{g}}$ as the sum of the series capacitances of SrTiO$_3$ and hBN assuming plane capacitors : $\frac{1}{C_{\text{g}}}=\frac{1}{C_{\text{STO}}}+\frac{1}{C_\text{BN}}\Rightarrow \frac{d_{\text{STO}}+d_\text{BN}}{\epsilon_\text{r}}=\frac{d_{\text{STO}}}{\epsilon_\text{STO}}+\frac{d_\text{BN}}{\epsilon_\text{BN}}.$ Since $d_\text{BN}\sim 10\:\text{nm}\ll d_{\text{STO}}=500\:\mu$m, we write the gate dielectric constant as :
\begin{equation}
\epsilon_\text{r}=\epsilon_\text{STO}\left(1+\frac{d_\text{BN}}{d_{\text{STO}}}\frac{\epsilon_\text{STO}}{\epsilon_\text{BN}}\right)^{-1}
\end{equation}
Numerically, $\epsilon_\text{BN}\simeq 3.6$ and $\epsilon_\text{STO}\sim 10^4$ at low temperature \cite{Sakudo1971}, so that $d_\text{BN}\epsilon_\text{STO}/d_{\text{STO}}\epsilon_\text{BN}\sim 0.1$. We can thus assume that $\epsilon_\text{r}\simeq\epsilon_\text{STO}$.\\

In order to estimate $\epsilon_\text{r}$ as a function of the back-gate voltage, $V_\text{g}$, we consider the tunneling conductance gate map of Fig. 2a (sample STO07) from which we can extract some values of the back-gate voltage at specific filling factors $\nu$. Note that the electron-hole asymmetry visible in this gate map stems from the non-linear behavior of $\epsilon_\text{STO}$ with gate voltage \cite{Hemberger1995,Sachs2014,Chen2020}. We plot in Fig. \ref{ExDFig1}a the line cut of the gate map at zero bias, averaged on a range of $\pm 20\:$meV around this value. We clearly observe the different non-zero conductance plateaus forming when $E_\text{F}$ is pinned inside one LL, with gaps in-between. As the gate voltages at the middle of those gaps correspond to completely filled and empty LLs, we identify the positions in gate voltage of the integer filling factors $\nu=-10,-6,-2,2$. Those values are converted into charge carrier density values $n$ in Fig. \ref{ExDFig1}b using $\nu=n\phi_0/B$ with $\phi_0=h/e$ the flux quantum and $n$ the charge carrier density. A polynomial of degree 5 fits and interpolates the evolution of $n$ with $V_\text{g}$. From this fit, charge neutrality at $n=0$ is achieved at $V^{\text{CNP}}=13.5\:$V. We then straightforwardly obtain the $V_\text{g}$-dependence of $\epsilon_\text{r}$ via :
\begin{equation}
\epsilon_\text{r}=\frac{d_{\text{STO}}}{\epsilon_0}\frac{en}{V_\text{g}-V^{\text{CNP}}}
\end{equation}

The red curve in Fig. \ref{ExDFig1}b shows the resulting $\epsilon_\text{r}$, which decreases with increasing gate voltage and ranges between $12\:500$ and $3\:500$. A similar $\epsilon_\text{r}(V_\text{g})$ profile but with slightly weaker values is obtained for sample AC23 ($3\:000<\epsilon_\text{r}<11\:500$). Finally, using the fit of the filling factor $\nu$, we can rescale the gate map as a function of $\nu$ as shown in Fig. \ref{ExDFig1}c. In particular, note that in the rescaled map the interaction-induced gap is maximal at charge neutrality $\nu=0$, as expected considering that the exchange interaction is maximal at half-filling of the zeroth Landau level.


\section{Landau levels of Dirac fermions}

\begin{figure}[ht!]
	\centering
		\includegraphics[width=1\textwidth]{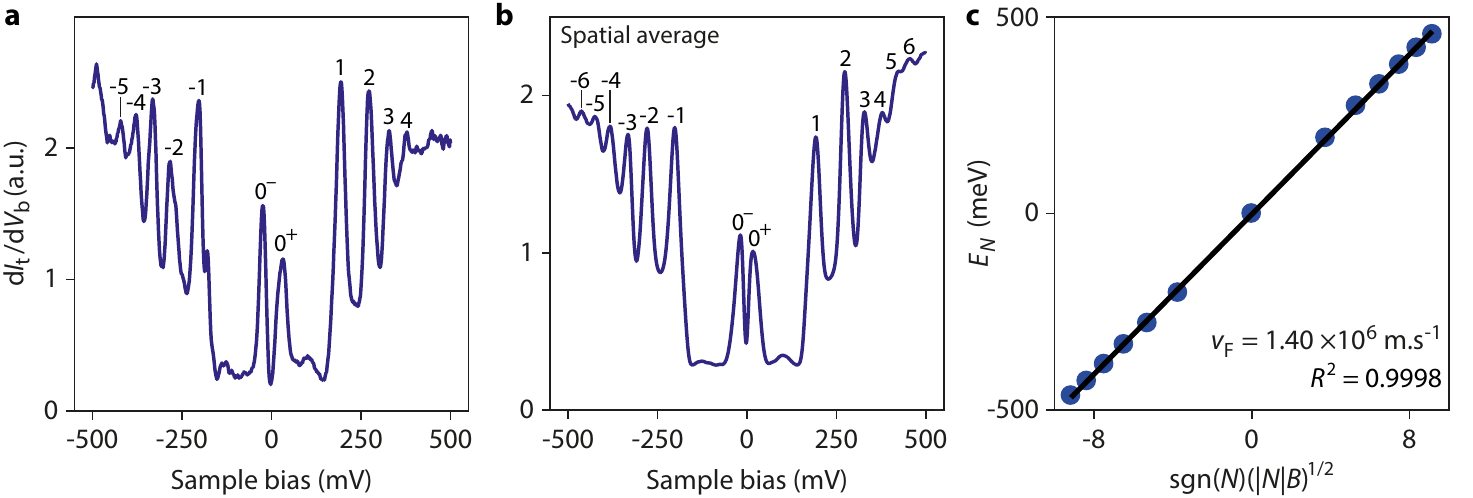}
	\caption{\textbf{Relativistic Landau levels in graphene in sample STO07 on SrTiO$\mathbf{_3}$.} \textbf{a}, Individual tunneling differential conductance d$\It$/d$\Vb$ spectrum at $B=14\:$T and $\Vg=13\:$V showing well-defined Landau level peaks. The zeroth Landau level is split into two peaks LL$_0{^\pm}$. \textbf{b}, Spatially-averaged d$\It$/d$\Vb$ spectrum under the same conditions on a $100\times 100\:$nm$^2$ area. \textbf{c}, Fit using Equation \eqref{LL_eq} of the positions of Landau levels $E_N$ as a function of $(|N|B)^{1/2}$ with $N$ the Landau level index. An excellent agreement is obtained and the fit yields a Fermi velocity $\vf=(1.403\pm 0.005)\times 10^6\:$m.s$^{-1}$.}
	\label{figS2}
\end{figure}

We show in Fig. \ref{figS2}a a tunneling conductance d$\It$/d$\Vb$ spectrum taken on sample STO07 (SrTiO$_3$) at $B=14\:$T (same than Fig. 2b of the main text), where graphene was brought to charge neutrality with a gate voltage $\Vg=13\:$V using the d$\It$/d$\Vb$ gate maps from Fig. 2a. Additionally, Fig. \ref{figS2}b displays a spatially-averaged d$\It$/d$\Vb$ spectrum on a $100\times 100\:$nm$^2$ area around the same position as Fig. \ref{figS2}a, where we clearly see well-resolved Landau levels in the local density of states up to $N=\pm 6$ as well as both peaks of the LL$_0$ broken-symmetry state, with a gap of $\Delta E^{0}\simeq 36\:$meV. We fit in Fig. \ref{figS2}c the positions $E_N$ of the Landau levels as a function of $\left(|N|B\right)^{1/2}$ ($E_0\approx 0.7\:$meV is taken at the middle of the LL$_0$ peaks). We obtain an excellent agreement with the theoretical dispersion relation for graphene :
\begin{equation}
E_N=\ed+\text{sign}(N)\vf\sqrt{2\hbar e|N|B}
\label{LL_eq}
\end{equation}
confirming the massless behavior of charge carriers in graphene. The fit yields a Fermi velocity of $\vf=(1.403\pm 0.005)\times 10^6\:$m.s$^{-1}$. This value is much greater than the expected theoretical one of $1.0\times 10^6\:\text{m.s}^{-1}$, which we attribute to the enhancement of electron-electron interactions at charge neutrality \cite{DasSarma2007,Luican2011,Chae2012}.\\

Interestingly, we note that this energy spacing of Landau levels is symmetric with respect to electron and hole levels. This indicates that a possible tip-induced local doping due to the work-function difference between the tip and graphene does not yield a significant spatial confinement on the Landau levels, which could have modified the Landau level spectrum~\cite{Ren21}.


\section{Tip-induced gating}

\begin{figure}[ht!]
	\centering
		\includegraphics[width=0.9\textwidth]{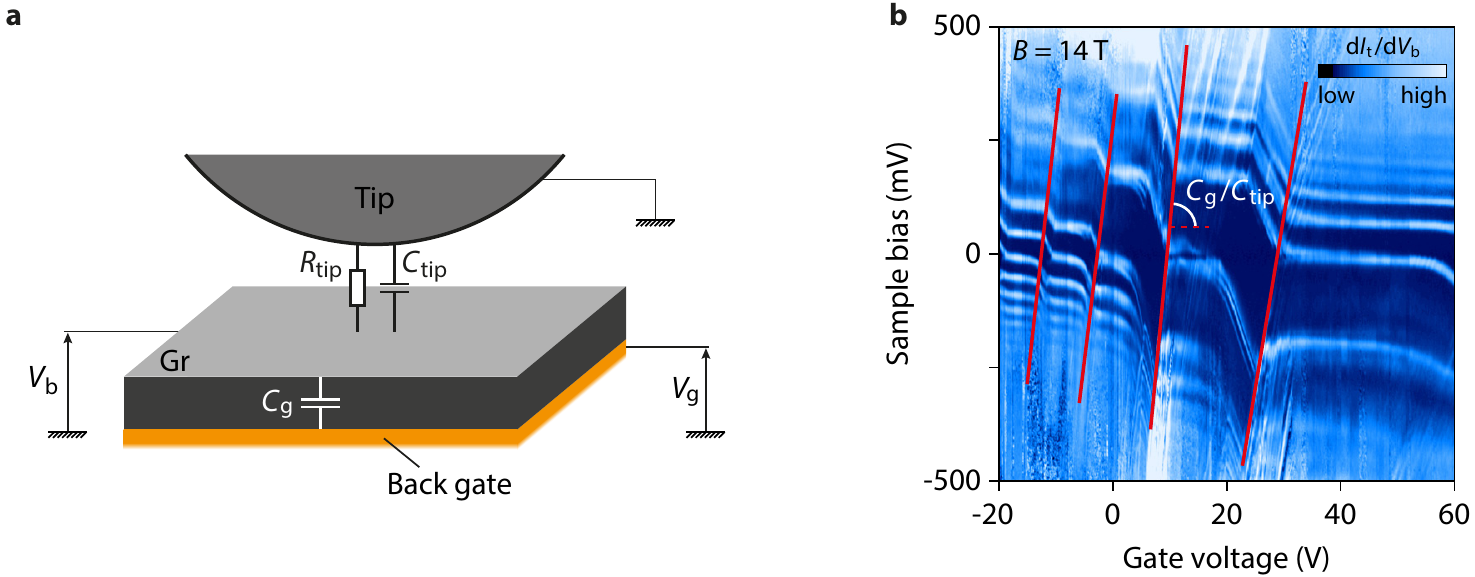}
	\caption{\textbf{Tip-induced gating.} \textbf{a}, Gating of graphene from both the back gate and the tip. \textbf{b}, Tunneling conductance gate map, same as Fig. 2a from the main text. The isodensity lines, represented in red, are tilted and their slope gives an estimation of the capacitances ratio $\Cg/\Ctip$.}
	\label{tip-gating}
\end{figure}

Probing graphene density of states in tunneling spectroscopy induces an unintentional local doping of graphene due to the field effect of the tip. The total doping of graphene from field effects of both the back gate and the tip is therefore :
\begin{equation}
n=\frac{1}{e}\left(\Cg\Vg-\Ctip\Vb\right)
\end{equation}
with the capacitances $\Cg$ between graphene and the back gate and $\Ctip$ between the tip and graphene, and the voltages $\Vg$ applied to the back gate and $\Vb$ applied to the sample (note that $\Vb\ll\Vg$) while the tip is grounded, as shown in Fig. \ref{tip-gating}a.\\

Constant-carrier-density lines in tunneling conductance gate maps thus do not appear as vertical lines at constant back-gate voltages, but rather as tilted lines, see the red lines in Fig. \ref{tip-gating}b which show that the Landau levels staircase transitions do not occur at the same back-gate voltage. The slopes of the isodensity lines are given by :
\begin{equation}
\frac{\diff n}{\diff\Vg}=0\Rightarrow \frac{\diff\Vb}{\diff\Vg}=\frac{\Cg}{\Ctip}
\end{equation}

Since the back-gate capacitance (per unit area) $\Cg$ can be estimated as $\Cg=\epsilon_0\epsilonr/d$, with $d$ the thickness of the back-gate insulator and $\epsilonr$ its relative dielectric constant, and the slope $\Cg/\Ctip$ of the isodensity lines can be obtained from the gate maps, following Ref. \cite{Chae2012}, it is possible to deduce the capacitance $\Ctip$. The variation of the carrier density due to tip-induced gating at constant back-gate voltage is then :
\begin{equation}
\delta n_\text{tip}=\frac{1}{e}\Ctip\Vb
\end{equation}
and the resulting tip-induced variation of the filling factor is :
\begin{equation}
\delta\nu_\text{tip}=\frac{\delta n_\text{tip}\phi_0}{B}
\end{equation}
with $\phi_0$ the flux quantum and $B$ the magnetic field.

\subsection{SiO$\mathbf{_2}$ sample}

The back-gate insulator is here comprised of the SiO$_2$ layer and the hBN flake. Its thickness is $d=d_{\text{SiO}_2}+d_\text{BN}\simeq 300\:$nm (see Table~S1). Its dielectric constant is written as (see Section IV for a similar derivation in the case of SrTiO$_3$ samples) :
\begin{equation}
\epsilonr=\epsilon_{\text{SiO}_2}\left(1+\frac{d_\text{BN}}{d_{\text{SiO}_2}}\frac{\epsilon_{\text{SiO}_2}}{\epsilon_\text{BN}}\right)^{-1}
\end{equation}

For $\epsilon_{\text{SiO}_2}=3.9$ and $\epsilon_\text{BN}\simeq 3.6$, we obtain $\epsilonr\simeq 3.7$. The back-gate capacitance is thus :
\begin{equation}
\Cg=\frac{\epsilon_0\epsilonr}{d}\simeq 10.9\:\text{nF.cm}^{-2}
\end{equation}

From several tunneling conductance gate maps we compute the slopes of the isodensity lines around charge-neutrality to be of the order of 0.1 (the tip capacitance changes with the tip apex shape). We then obtain an estimation of the tip capacitance as :
\begin{equation}
\frac{\Cg}{\Ctip}\approx 0.1\Rightarrow \Ctip\approx 110\:\text{nF.cm}^{-2}
\end{equation}
~
We now focus on the tip-induced gating at $\nu=0$ at the bias positions of the two peaks of the split LL$_0$, positions that we used to estimate the $\nu=0$ gap $\Delta E^0(B)$. The tip-induced gating at those biases is given by :
\begin{equation}
\left|\delta n_\text{tip}\right|=\frac{\Ctip}{e}\frac{\Delta E^0(B)}{2}
\end{equation}
which yields for the tip-induced variation of the filling factor ($\nu=0$ at $\Vb=0$, set by the gate voltage) :
\begin{equation}
\left|\delta\nu_\text{tip}\right|=\frac{\phi_0\Ctip\Delta E^0(B)}{2eB}
\label{delta_nu_tip}
\end{equation}
~
Numerically, at $B=14\:$T, we measured $\Delta E^0\simeq 55\:$meV (see Fig. 2e from the main text), which gives $\left|\delta\nu_\text{tip}\right|\simeq 0.06$. At $B=1.5\:$T, we measured $\Delta E^0\simeq 15\:$meV, which gives $\left|\delta\nu_\text{tip}\right|\simeq 0.14$. In both cases, the tip-induced variation of the filling factor is negligible and we can assume it does not alter the physics of the $\nu=0$ state we are probing. Note also that since $\Delta E^0$ scales as the Coulomb energy $\mathcal{E}_\text{C}\propto\sqrt{B}$, this yields $\left|\delta\nu_\text{tip}\right|\propto B^{-1/2}$ and the effect of the tip on the filling factor decreases at higher magnetic fields.

\subsection{SrTiO$\mathbf{_3}$ samples}

The back-gate insulator is here comprised of the SrTiO$_3$ substrate and the hBN flake. Its thickness is $d\simeq d_\text{STO}=500\:\mu$m and its dielectric constant $\epsilonr$ can be determined as explained in Section IV. From Fig. \ref{ExDFig1}b we can estimate $\epsilonr\approx 7\:500$ at charge-neutrality for sample STO07. We then compute the back-gate capacitance as : 
\begin{equation}
\Cg=\frac{\epsilon_0\epsilonr}{d}\simeq 13.3\:\text{nF.cm}^{-2}
\end{equation}
~
We obtain from several tunneling conductance gate maps a similar ratio of the capacitances of the order of :
\begin{equation}
\frac{\Cg}{\Ctip}\approx 0.1\Rightarrow \Ctip\approx 133\:\text{nF.cm}^{-2}
\end{equation}
~
Using Equation \eqref{delta_nu_tip}, we estimate at $B=14\:$T where we measured $\Delta E^0\simeq 40\:$mV a tip-induced variation of the filling factor around $\nu=0$ of $\left|\delta\nu_\text{tip}\right|\simeq 0.05$. At $B=6\:$T where $\Delta E^0\simeq 15\:$meV, we have $\left|\delta\nu_\text{tip}\right|\simeq 0.04$. The effect of the tip on the filling factor is therefore also negligible.


\section{Contact potential difference measurements}

\begin{figure}[ht!]
	\centering
		\includegraphics[width=1\textwidth]{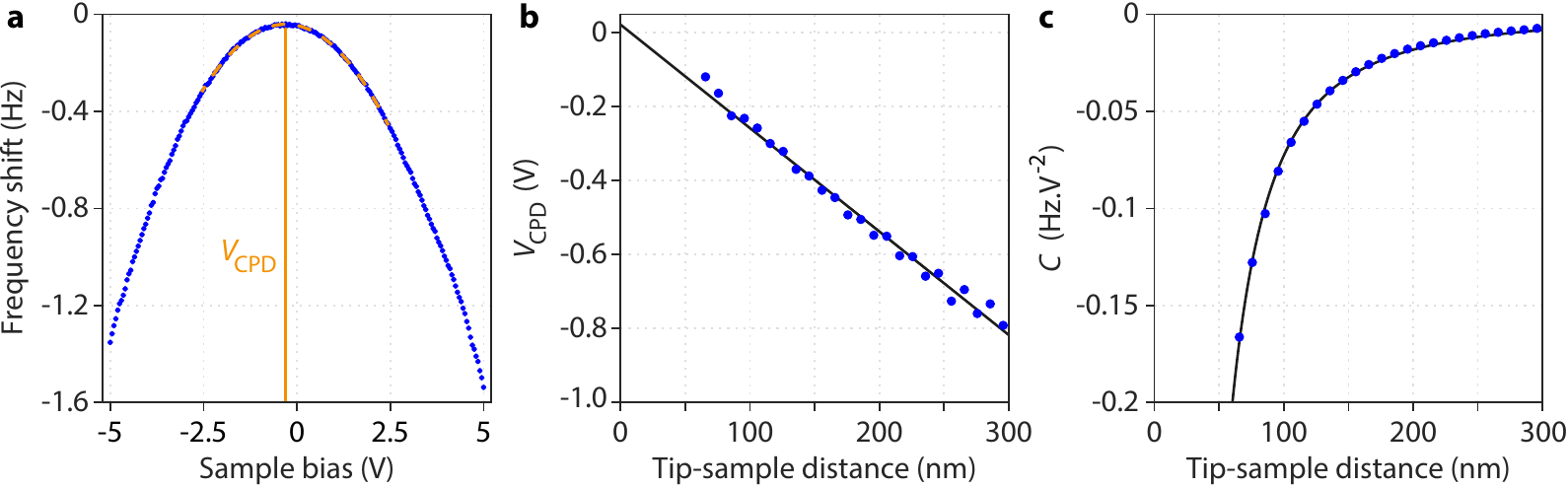}
		\vspace{-2.5em}
	\caption{\textbf{Contact potential difference measurements on sample AC04.} \textbf{a,} Frequency shift $\Delta f$ as a function of sample bias $\Vb$ (blue points) at a tip-sample distance of $z=115\:$nm from graphene. $\VCPD(z)$ is determined as the maximum of the fitting parabola in dashed yellow. \textbf{b,} $\VCPD$ as a function of tip-sample distance $z$ from graphene (blue points). The CPD between the tip and graphene is determined by the intercept of the linear fit (black line) with $z=0$. \textbf{c,} Evolution of the curvature $C$ of the parabolas as a function of the tip-sample distance $z$ from graphene (blue points). The fit using Equation \eqref{fitting_curvature} (black line) yields the tip-sample contact position $z_0$, from which we compute the tip-sample distance $z=z_\text{scan}-z_0$.}
	\label{fig_S4}
\end{figure}

We show here contact potential difference $\VCPD$ measurements performed in AFM mode on the graphene as a function of the tip-sample distance, which enables us to extract the exact band-bending beneath the tip due to the work-function difference between the tip and graphene. We start from the contact in AFM mode and then retract the tip at a position $z_\text{scan}$. The frequency shift $\Delta f$ of the tuning fork is measured as the sample voltage $\Vb$ is swept from $-5\:$V to $+5\:$V while the tip is kept grounded. We obtain a parabola described by the following equation :
\begin{equation}
\Delta f=C\left(\Vb-\VCPD\right)+\Delta f_\text{min}
\label{EFM_parabola}
\end{equation}
such as the one shown in Fig. \ref{fig_S4}a. $\VCPD$ is estimated as the position of the maximum of the parabola, using a quadratic function to fit the curve. The variation of $\VCPD$ with the tip-sample distance is shown in Fig. \ref{fig_S4}b. A linear dependence is observed, and the intrinsic $\VCPD_0$ between the tip and graphene can be determined by the intercept of the linear fit with $z=0$, where $z$ is the absolute distance between the tip and graphene, given by $z=z_\text{scan}-z_0$, with $z_0$ the tip-sample contact position.\\

In order to compute $z_0$, we fit in Fig. \ref{fig_S4}c the evolution of the curvature coefficient $C$ of the parabola with the tip-sample distance using the following equation, see Ref. \cite{Laurent2012} :
\begin{equation}
C=\frac{\beta}{\left(z_\text{scan}-z_0\right)^2}
\label{fitting_curvature}
\end{equation}
with $z_0$ and $\beta$, the force probe calibration parameter, as the two fitting parameters. The obtained value of $z_0$ enables us to plot Figs. \ref{fig_S4}b and c as a function of the absolute tip-sample distance $z$. From Fig. \ref{fig_S4}b we obtain a negligible contact potential difference of $\VCPD_0=0.02\:$V for a tip-sample distance in tunnel mode, that is, of the order of $1\:$nm. This very small potential can be accounted for by the fact that the tip apex is covered with gold due to the tip reshaping on the gold contact, and is not made of PtIr unlike the bulk of the wire. The work-function of gold is about $4.8\:$eV \cite{Anderson1959} while graphene is given at $4.6\:$eV \cite{Yan2012}. Our measurements show that these two work-functions for our tip apex and graphene on hBN are actually very close and thus do not induce a significant band-bending. This must be compared to the work-function of PtIr which is $5.7\:$eV \cite{Moran2016}, which would lead to a much more unfavorable situation with a significant band-bending.


\section{Tunneling conductance gate maps and interaction-induced gap in sample AC04}

\begin{figure}[ht!]
	\centering
		\includegraphics[width=1\textwidth]{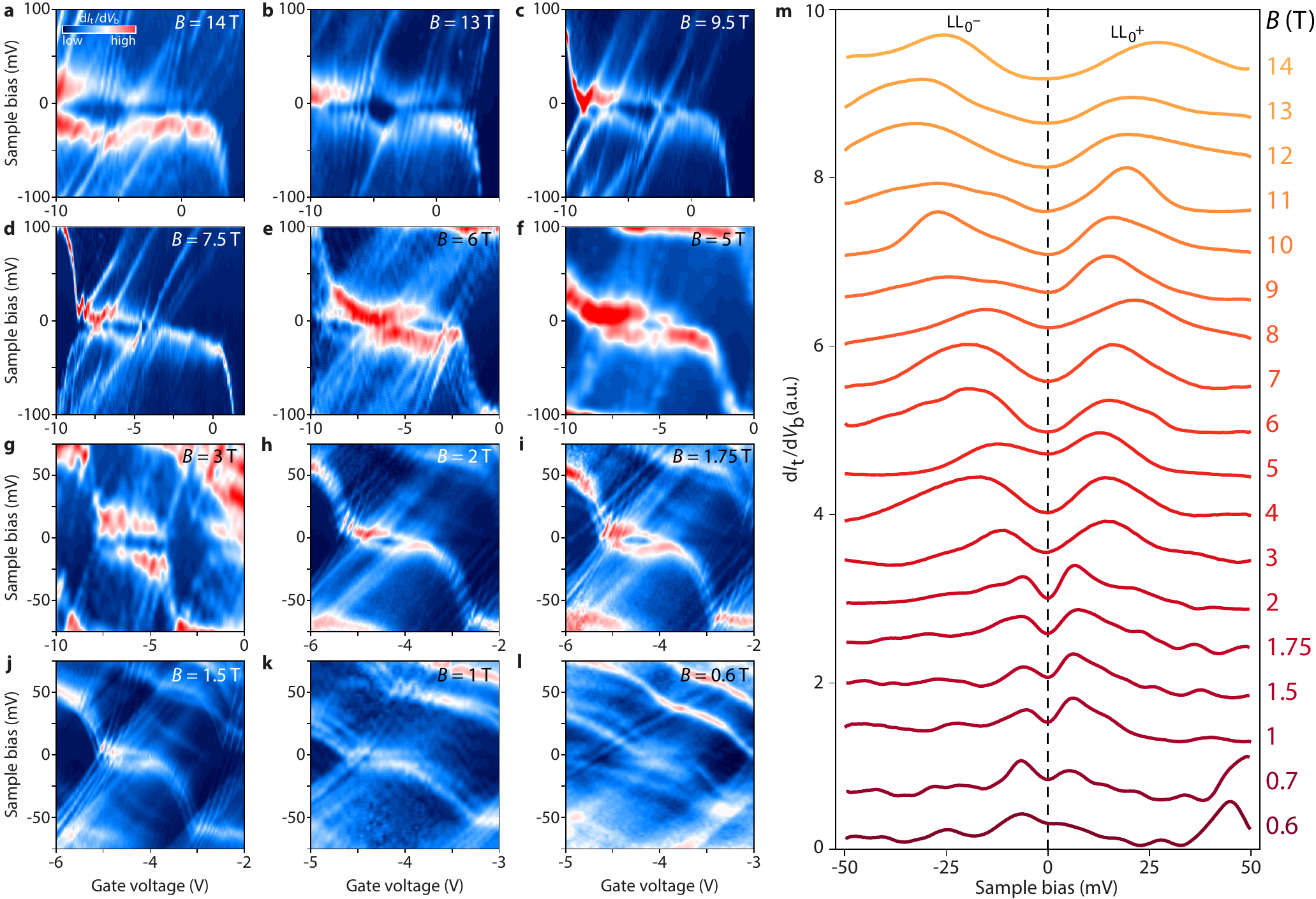}
	\caption{\textbf{Evolution of $\bm{\Delta E^{0}}$ with the magnetic field on hBN/SiO$_{\bm{2}}$,} in sample AC04. \textbf{a-l}, Tunneling conductance, d$\It$/d$\Vb$, gate maps at different magnetic fields. \textbf{b,c} were performed at the same position, same for \textbf{h,l}. \textbf{m}, Some d$\It$/d$\Vb$ spectra, centered around zero bias, used to estimate $\Delta E^{0}$ from d$\It$/d$\Vb$ gate maps at different magnetic fields $B$.}
	\label{figS5}
\end{figure}

We show in Fig. \ref{figS5} some examples of $\Delta E^{0}$ estimations at different magnetic fields $B$ for the sample AC04 (Gr/hBN/SiO$_2$) which were reported in Fig. 2e of the main text. $\Delta E^{0}$ is computed as the maximum value of the peak-to-peak energy between LL$_{0^\pm}$ extracted from the individual spectra of the gate map.

\section{Tunneling conductance gate maps and interaction-induced gap in sample STO07}

\begin{figure}[p]
	\thisfloatpagestyle{empty}
	\centering
	\vspace{-3.2em}
		\includegraphics[width=0.99\textwidth]{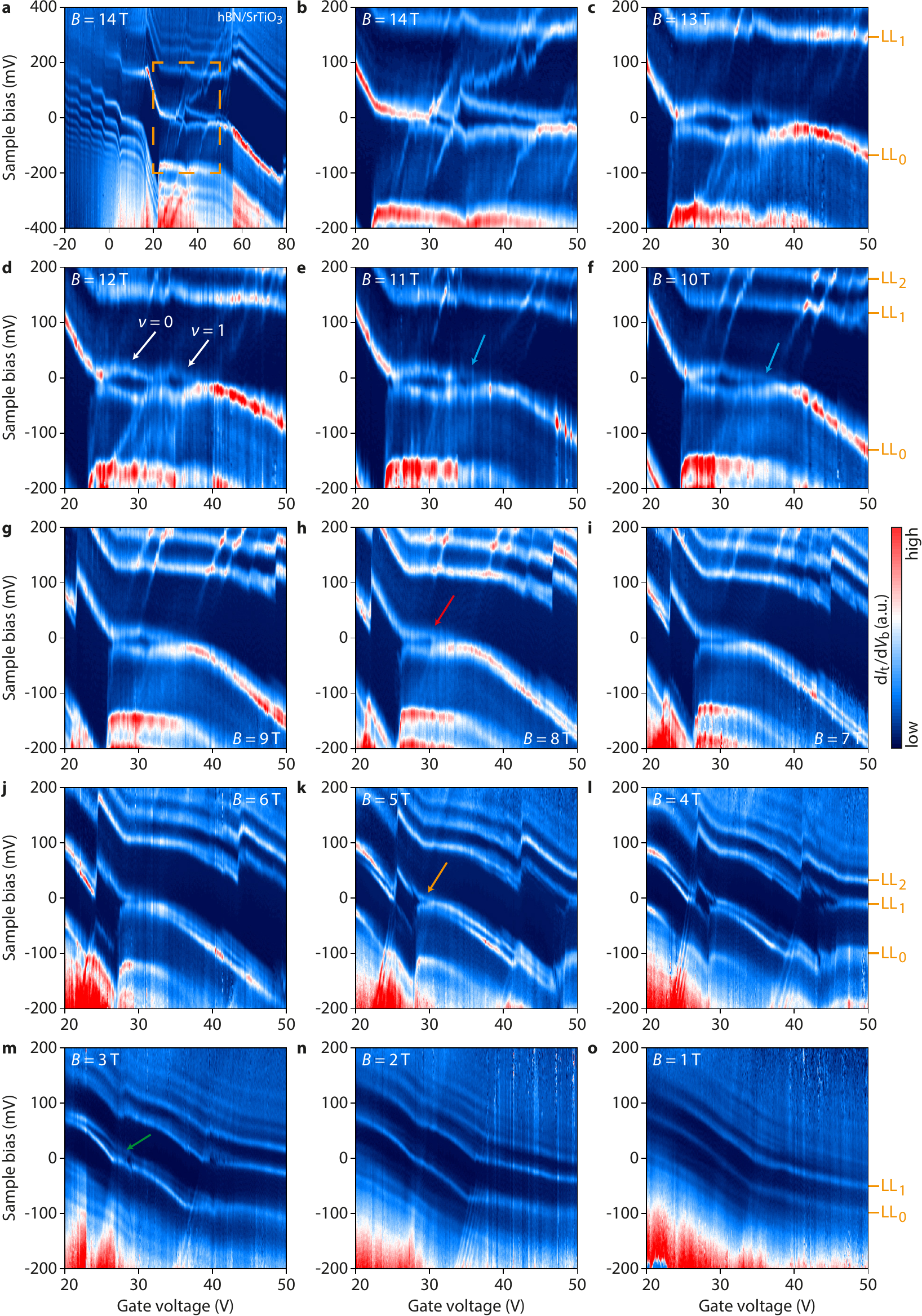}
	\caption{\textbf{Tunneling conductance gate maps as a function of the magnetic field $\bm{B}$ on hBN/SrTiO$_\mathbf{3}$,} in sample STO07 and at the same position.}
	\label{figS6}
\end{figure}

Figure \ref{figS6} shows the evolution of the tunneling conductance d$\It$/d$\Vb$ gate maps for decreasing magnetic fields from $B=14\:$T to $B=1\:$T in sample STO07 (SrTiO$_3$ substrate). For each magnetic field, we perform the same gate sweep from $\Vg=80\:$V to $\Vg=-20\:$V in order to keep the hysteresis cycle of the SrTiO$_3$ substrate constant. The full d$\It$/d$\Vb$ gate map at $B=14\:$T is shown in panel (a), and the zoom on the yellow rectangle centered on the LL$_0$ gap is displayed in panel (b). The next panels are zooms on the same area as (b). All those d$\It$/d$\Vb$ gate maps were acquired at the same position on graphene, up to the magnetic field drift.\\

Let us start at $B=1\:$T in the bottom right panel. Many peaks dispersing negatively with the gate voltage are visible, among them we can already distinguish LL$_0$ and LL$_1$. With increasing magnetic fields, other resonant peaks eventually merge with each other to form LLs, as described in Ref. \cite{Gutierrez2018}. For instance, LL$_2$ is formed at $B=3\:$T while LL$_{-1}$ becomes distinguishable at $B=5\:$T. We also notice at $B=1\:$T that there is no pinning effect of $\ef$ inside LLs, which thus disperse continuously with the gate voltage. LLs start to pin the Fermi level at $B=3\:$T with the formation of a small plateau for LL$_0$ at $\Vg=27.5\:$V. However note that the splitting of LL$_0$ in panel (m) is mostly due to the lifting of the orbital degeneracy \cite{Luican-mayer2014}, such that the apparent gap at zero bias (indicated by the green arrow) may be different from the interaction-induced gap we are aiming for. This orbital splitting is maximum at $B=4\:$T and then decreases at higher magnetic fields.\\

The gap $\Delta E^{0}$ finally opens at $B=5\:$T, see the yellow arrow in panel (k). Since the density of states of LL$_0$ grows with $B$, the Fermi level stays pinned inside LL$_0$ for a wider range of gate voltage with increasing $B$, and as a result the gap develops on a larger LL$_0$ plateau. At $B=8\:$T the gap features a maximum due to its enhancement by exchange interactions, see the red arrow in panel (h). At $B=10\:$T, a second maximum appears  on the right extremity of the LL$_0$ plateau, see the blue arrow in panel (f), while at $B=12\:$T we clearly distinguish two lobes marked by the white arrows in panel (d). The left lobe corresponds to the opening of the $\nu=0$ gap whereas the right one is due to the opening of the $\nu=1$ gap (however the $\nu=-1$ gap is not visible).


\section{Asymmetry of the KB pattern.}

\begin{figure*}[!ht]
	\centering
	\includegraphics[width=1\linewidth]{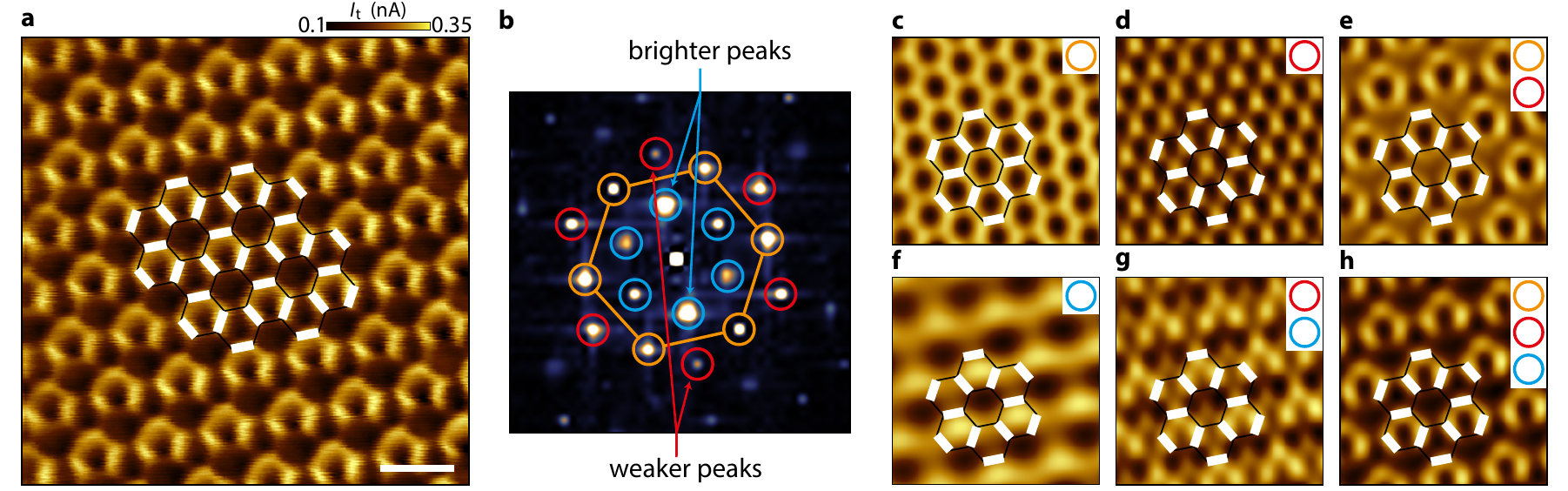}
	\caption{\textbf{2D-FT decomposition of the asymmetric Kekul\'e distortion.} \textbf{a}, $3\times 3\:$nm$^2$ image showing an asymmetric KB pattern, measured at $B=14\:$T and $V_\text{b}=2\:$mV. \textbf{b}, 2D Fourier transform (2D-FT) of the STM image in \textbf{a}.  Yellow circles indicate peaks of the honeycomb lattice, red and blue circles indicate peaks of the bond-density wave. \textbf{c-h}, Filtered images obtained by considering certain peaks of the FFT as indicated in the top right corner of each panel. The Kekul\'e lattice is drawn in white for reference. The KB order is mostly retrieved by considering only the yellow and red peaks. The asymmetry of the KB pattern is encoded in the blue peaks whose two of them are twice as high as the others due to the K-CDW order. The scale bar is the same for all figures : $500\:$pm.}
	\label{ExDFig2}
\end{figure*}

We show in Fig.~\ref{ExDFig2}a a STM image of an asymmetric KB pattern. Fig.~\ref{ExDFig2}b displays the corresponding 2D Fourier transform (2D-FT). 
The 2D-FT is mainly comprised of three hexagons, defined by the yellow, red and blue encircled peaks. To decipher the 2D-FT, we filter the STM image by considering certain peaks only. The yellow peaks alone yield the usual honeycomb lattice in Fig.~\ref{ExDFig2}c. The red peaks give the image shown in Fig.~\ref{ExDFig2}d, which features a triangular lattice. When we superimpose the KB lattice drawing, we notice that each bright point of the triangular lattice in Fig.~\ref{ExDFig2}d falls either on the strong white bonds of the Kekul\'e lattice or at the center of the hexagons devoid of strong bond: the addition of both images yields the bond-density wave shown in Fig.~\ref{ExDFig2}e where we have filtered the STM image by considering this time both yellow and red peaks and mostly recovered the original KB pattern. This also justifies why the hexagon devoid of strong bond in the KB pattern appears brighter than the neighboring hexagons comprised of three strong bonds, similarly visible in Fig.~4a of the main text. Note that the presence of two red peaks with halved amplitude in one direction is responsible of the slight asymmetry that is already visible in Fig.~\ref{ExDFig2}e.

We show in Fig.~\ref{ExDFig2}f the image obtained after filtering using only the blue peaks. We observe a strongly asymmetric triangular lattice encoding the Kekul\'e spatial modulation at $\sqrt{3}$ times the graphene lattice parameter. The asymmetry arises from a large asymmetry between the blue peaks in the 2D-FT, where two peaks in one direction are twice as high as the others. This yields dissimilar weights to the bond-density wave, as shown in Fig.~\ref{ExDFig2}g where we have filtered considering red and blue peaks, and explains the strong asymmetry we observe in the KB pattern, which is fully recovered in Fig.~\ref{ExDFig2}h where we have filtered with the yellow, red and blue peaks. We conjecture that this strong asymmetry of the 2D-FT originates from the existence of the K-CDW order whose contribution is visible in Fig.~\ref{ExDFig2}f, since in symmetric KB pattern (where this K-CDW order is not visible) there is no such asymmetry between the blue and red peaks.

\section{Bias dependence of the KB order.}

\begin{figure*}[ht!]
	\centering
	\includegraphics[width=0.7\linewidth]{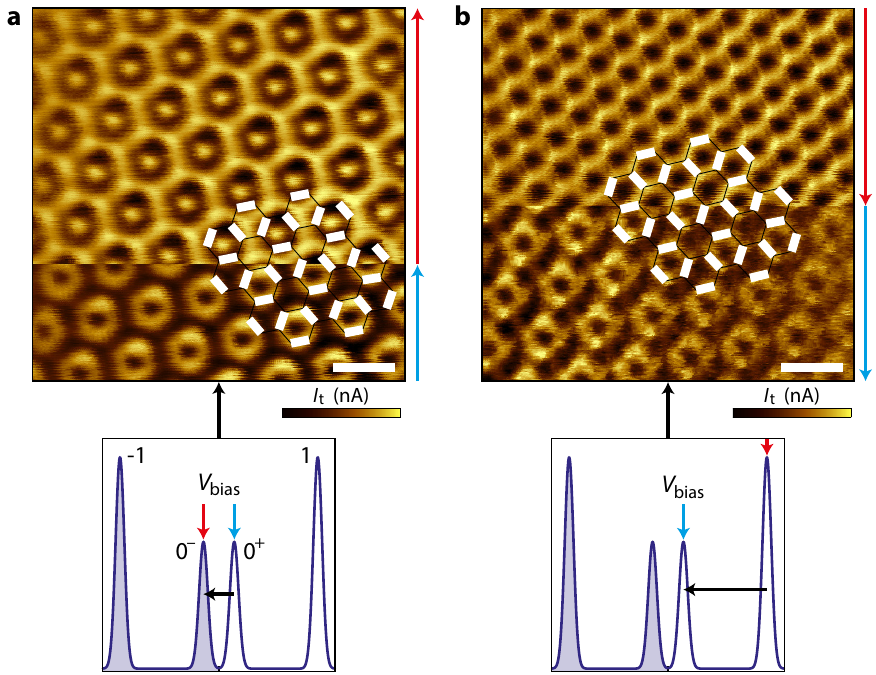}
	\caption{\textbf{Contrast inversion and emergence of the Kekul\'e bond order.} $3\times 3\:$nm$^2$ STM images during which we changed the bias voltage as shown in the bottom insets (the current color bars are tuned separately for each half of the images). \textbf{a}, We start (bottom) at $V_\text{b}=32\:$mV (LL$_{0^+}$) and switch (top) to $V_\text{b}=-12\:$mV (LL$_{0^-}$) to observe a contrast inversion of the KB lattice. \textbf{b}, We start (top) at $V_\text{b}=200\:$mV (LL$_1$) and switch (bottom) to $V_\text{b}=20\:$mV (LL$_{0^+}$) and observe the emergence of the KB order from the honeycomb lattice. Scale bars for both images are $500\:$pm.}
	\label{ExDFig3}
\end{figure*}

We show in Fig. \ref{ExDFig3} constant-height mode STM images where we have changed the sample bias $V_\text{b}$ during scanning. The red and blue arrows on the right of each image show the direction of the slow scan axis, and their color corresponds to the actual sample bias which is indicated in the bottom figures. In Fig. \ref{ExDFig3}a we clearly observe a contrast inversion when switching the sample bias from LL$_{0^+}$ to LL$_{0^-}$, with the continuity of the KB pattern at the interface. In Fig. \ref{ExDFig3}b we see the transition from the usual honeycomb lattice to the KB pattern when switching the sample bias from LL$_1$ to LL$_{0^+}$. This indicates that both the KB order and the underlying K-CDW disappear when the bias voltage is away from the zLL peaks, which rules out a tip-induced artifact as the origin of the observed KB order.

\section{K-CDW configuration change.}

\begin{figure*}[!b]
	\centering
	\includegraphics[width=1\linewidth]{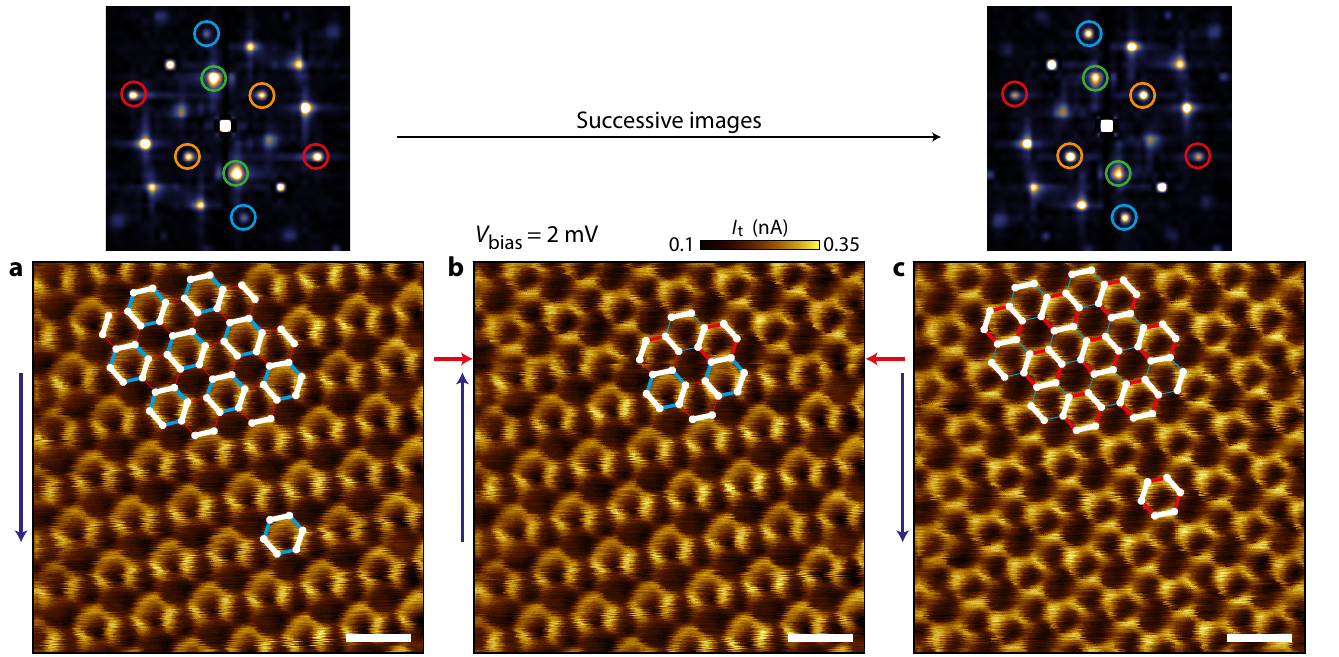}
	\caption{\textbf{Asymmetry reversal of the Kekul\'e pattern.} $3\times 3\:$nm$^2$ STM images measured at $B=14\:$T, $V_\text{b}=2\:$mV and at the same position. The three images were measured successively (scanning time : 1 min). A jump occurs in \textbf{b} at the scan line indicated by the red arrows, leading to an inversion of the asymmetry of the Kekul\'e pattern. The slow scan axis direction is indicated by the blue arrows on the left of each image. Scale bars for the three images are $500\:$pm.}
	\label{ExDFig4}
\end{figure*}

We illustrate here the time-varying nature of the K-CDW. We show in Fig. \ref{ExDFig4} three successive images acquired in a row at the same position and $V_\text{b}=2\:$mV. The vertical arrow on the left of each image indicates the direction of the slow scan axis. Figure~4d belongs to the same set of image acquisition. The lattice in overlay describes the asymmetric KB pattern, with the white links being the strong bonds of the KB order, whereas the asymmetry that comes from the K-CDW order makes the hexagons with blue weak bonds brighter than the hexagons with red weak bonds. The next image in panel (b) (duration of each image : 53 seconds) starts from the bottom, where we observe the same KB pattern. However, a jump occurs at the line indicated by the red arrows, and, after that, in the top part of the image, the asymmetry of the KB pattern is reversed. Using the lattice in overlay as a guide for the eye, we see that the red hexagons are brighter (due to the three strong white bonds almost merging together), such that the new pattern is the mirror of the previous one. Eventually, the next image in panel (c) displays this new pattern with brighter red hexagons on the whole area, and the next images we realized during several minutes happened to be identical. This indicates that the K-CDW order transited and reversed the asymmetry of the KB phase. Note that the pattern of the strong white bonds, which defines the KB order, stays unchanged in the three images (in opposition to the KB order transition shown in Figs. 4f and g).\\

\begin{figure*}[!b]
	\centering
	\includegraphics[width=0.8\linewidth]{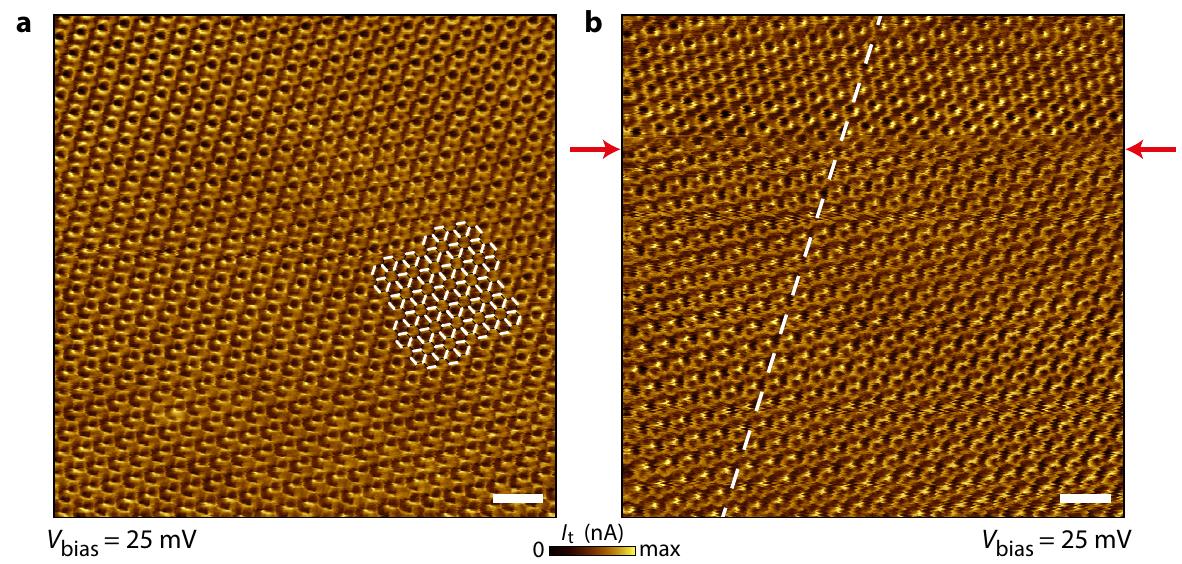}
	\caption{\textbf{Change of the Kekul\'e asymmetry.} $10\times 10\:$nm$^2$ STM images measured at $B=14\:$T and $V_\text{b}=25\:$mV. In \textbf{b}, the asymmetry pattern changes at the scan line indicated by the red arrows. Scale bars for both images are $1\:$nm.}
	\label{ExDFig5}
\end{figure*}

The asymmetry reversal of the KB order due to the K-CDW transition is well seen in the 2D-FT of both images in Figs. \ref{ExDFig4}a and c, see top insets. For Fig. \ref{ExDFig4}a, the K-CDW appears in the inner hexagon, where the two peaks encircled in green are twice as bright as the other four peaks. On the contrary, for Fig. \ref{ExDFig4}c, this is now the yellow peaks that are brighter than the other four, with the amplitude of the green peaks lowered. The change of the direction of the two brighter peaks induces the change of the asymmetry pattern of the KB order. Interestingly, the outermost hexagon, which corresponds to the bond-density wave, also features a change in the intensity of its peaks : in Fig. \ref{ExDFig4}a the blue peaks are halved in amplitude while in Fig. \ref{ExDFig4}c the red peaks are halved. This does not yield any significant change of the KB pattern but this may mean that the bond-density wave and the K-CDW are entangled.\\

Therefore, the asymmetry of the KB patterns we observed depends on the K-CDW order and its fluctuations. Moreover, Fig. \ref{ExDFig4}b shows an image in which the K-CDW switched from that of Fig. \ref{ExDFig4}a to that of Fig. \ref{ExDFig4}c during the acquisition. This change that occurred during the scan indicates that the K-CDW switches instantaneously on the time-scale of the scan speed. This can also indicates either a change of the K-CDW on the entire sample in case of a homogeneous K-CDW, or the displacement of domains with different K-CDW configurations separated by domain walls. Note that this only concerns this anomalous K-CDW, which coexists with the KB order, the latter being unchanged in the three images (the bright bonds pattern remains the same).\\

Such transitions of the K-CDW happened a few times during our measurements. In Fig. \ref{ExDFig5}a we show a $10\times 10\:$nm$^2$ image of an asymmetric KB pattern with the circle-like pattern formed by the merging of the strong bonds inside one hexagon of the KB order unit cell. Imaging the same area a few minutes later in Fig. \ref{ExDFig5}b unveils a spontaneous change of K-CDW configuration, similar to that in Fig. \ref{ExDFig4}b, which occurred during scanning on the line indicated by the red arrows: in the top part, the three strong bonds merge together inside another hexagon of the KB order unit cell with respect to the bottom part (see the white dashed line which intercepts the circles in the top part of the image and, conversely, passes between the circles in the bottom part). As previously, the KB order lattice itself does not change.\\

We point out that we cannot exclude a K-CDW configuration change induced by the action of the scanning tip. Still, such a tip-induced change also implies that the K-CDW is not pinned and can be subject to fluctuations.


\section{Induced $\mathbf{t_2}$ asymmetry in the charge-density-wave state.}

\begin{figure*}[ht!]
\includegraphics[width=\linewidth]{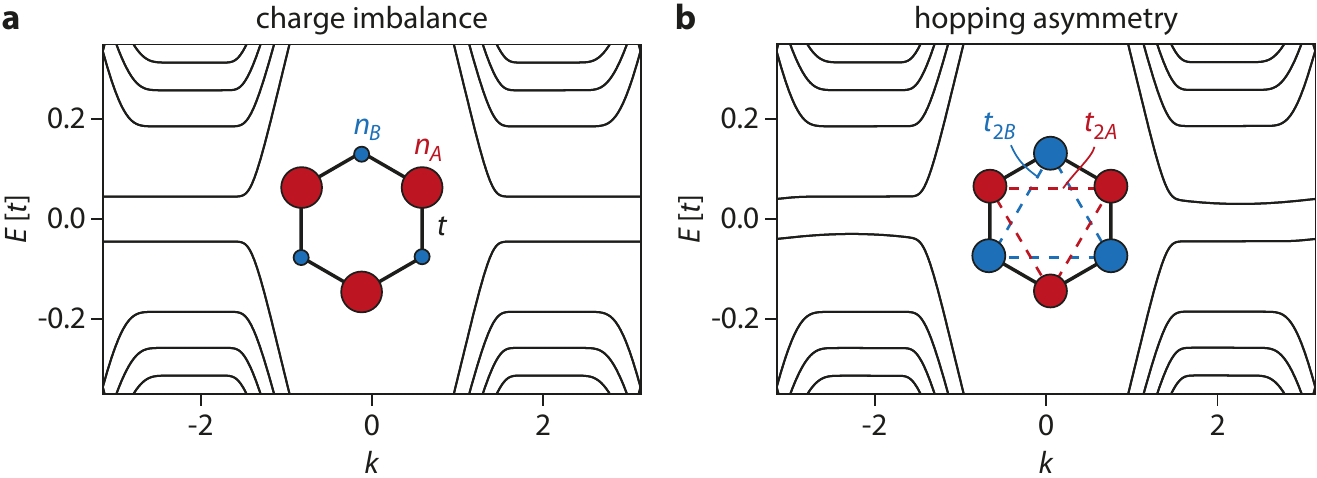}
\centering
\caption{\textbf{Effect of sublattice charge imbalance and a $\bm{t_2}$ asymmetry on the zeroth Landau level.} \textbf{a} shows that the effect of a finite charge imbalance $\Delta n = n_A-n_B$ is to gap the zeroth Landau Level of graphene. \textbf{b} shows that a hopping asymmetry $\Delta t_2 = t_{2,A}-t_{2,B}$ also opens up a gap, that depends on momentum $k$ as we move away from the $K$ and $K'$ points. The parameters are chosen so that $E_\text{g}$ is the same on both plots at the $K$ and $K'$ points, according to Equation \eqref{eq:gapzLL}. Simulations were performed using the \textsc{kwant} software \cite{Groth_2014} for a $41\times 41$ hexagonal lattice with $\phi=0.003$ flux per plaquette, in units of the flux quantum. 
Energies are measured in units of the nearest-neighbor hopping $t$. For \textbf{a}, $\Delta t_2 = 0$ and $\Delta n = 0.045$ , while for \textbf{b}, $\Delta t_2 = 0.015$ and $\Delta n=0$.}
\label{ExDFig6}
\end{figure*}

We discuss here how a second nearest-neighbour hopping asymmetry gaps the zeroth Landau level (zLL) of graphene. {We consider the spinor $\psi=(\psi_{AK},\psi_{BK},\psi_{AK'},\psi_{BK'})$, where $\psi_{\sigma \tau}$ is a zLL single-particle wavefunction in sublattice $\sigma$ and valley $\tau$. In this basis,} both the sublattice imbalance {$\Delta n=n_A-n_B$} and the second nearest-neighbor hopping asymmetry $\Delta t_2 = t_{2A}-t_{2B}$ (see Fig. \ref{ExDFig6}) enter the low-energy Hamiltonian close to the Dirac point with the matrix $\tau_0\otimes\sigma_z$ in valley ($\tau$) and  {sublattice} space ($\sigma$).
This matrix structure implies that both perturbations gap out the $K$ and $K'$ points of graphene, with a gap given by [\textcolor{blue}{30}] :
\begin{equation}
	\label{eq:gapzLL}
	E_\text{g} = \Delta n + \dfrac{3}{2}\Delta t_2,
\end{equation}
which is of equal sign for both valleys. We can visualize the effect of $\Delta t_2$ and $\Delta n$ on the zeroth Landau level by diagonalizing the graphene Hamiltonian in the presence of a magnetic field.
The spectrum with $\Delta t_2\neq 0$ and $\Delta n \neq 0$ are shown in Figs.~\ref{ExDFig6}a and b respectively, obtained with the \textsc{kwant} package \cite{Groth_2014}. We can confirm numerically that the gap is given by Equation \eqref{eq:gapzLL} and that when $\Delta n=-\dfrac{3}{2}\Delta t_2$ the gap closes, confirming that both perturbations enter the Hamiltonian with the same matrix structure.

\begin{figure*}[ht!]
\includegraphics[width=\linewidth]{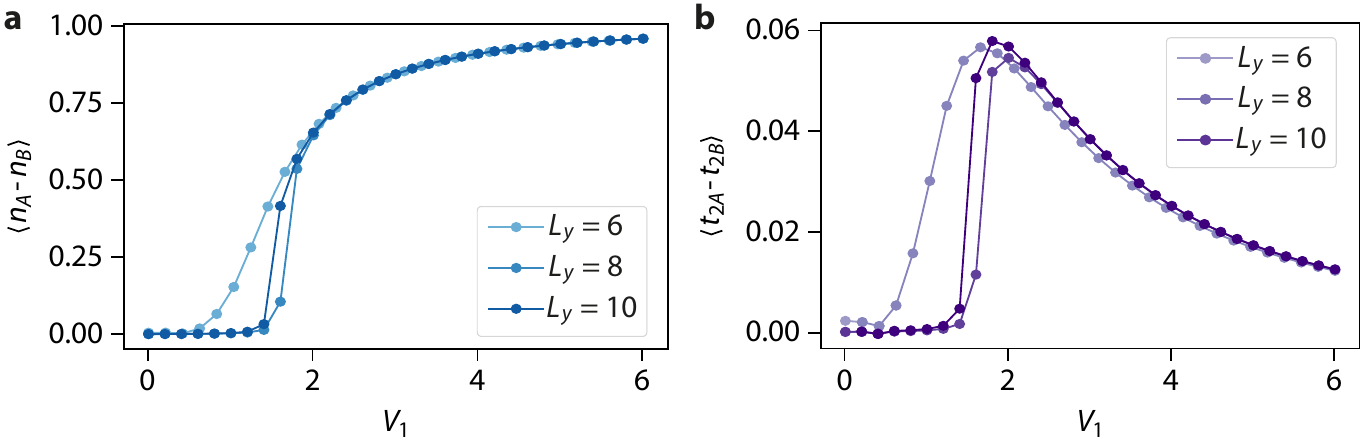}
\centering
\caption{\textbf{Induced $\bm{\Delta t_2}$ asymmetry by interactions.} \textbf{a} shows that a sublattice charge imbalance $\Delta n=\left\langle n_A-n_B\right\rangle\neq 0$ develops as $V_1$ increases. 
\textbf{b} shows the concomitant emergence of a second nearest-neighbor bond asymmetry $\Delta t_2=\left\langle t_{2A}-t_{2B}\right\rangle \neq 0$, peaking at intermediate values of $V_1$.
The simulations are carried out for cylinder circumferences of $L_y=6,8,10$ sites, all with bond-dimension $\chi=1000$, using the \textsc{tenpy} package \cite{Hauschild2018}.}
\label{ExDFig7}
\end{figure*}

The above argument suggests that interactions that induce a finite $\Delta n$ will generically induce a finite $\Delta t_2\neq 0$, as they both enter with the same matrix structure. To exemplify this generic behavior we use the Hamiltonian of graphene in the presence of nearest-neighbor interactions $V_1$ :
\begin{equation}
	\label{eq:graphenham}
	H = -t\sum_{\left\langle ij\right\rangle} \left( c^\dagger_i c_j + \mathrm{h.c.}\right)+ V_1 \sum_{\left\langle ij\right\rangle} n_i n_j,
\end{equation}
where the sums are taken over nearest neighbors of the honeycomb lattice. Note in particular that the second-nearest neighbor hopping is {explicitly} zero {in the Hamiltonian}. In the limit of infinitely large interaction, the ground state of $H$ at half-filling is a charge-density wave with one fully occupied and one fully empty {sublattice}, a state characterized by $\Delta n =1$. 
The bond asymmetry $\Delta t_2$ is expected to be exactly zero in this limit, since all sites on one sublattice are completely full and thus no states are available to hop to. Similarly,  all sites on the other sublattice are completely empty such that no states are available to hop from.
At sufficiently large {(but finite)} $V_1/t$ the ground state is a charge-density wave with partial sublattice imbalance, as we numerically show in Fig.~\ref{ExDFig7}. In Fig.~\ref{ExDFig7}a, we show the expectation value $\Delta n= \left\langle c^\dagger_{iA} c_{iA}-c^\dagger_{iB}c_{iB}\right\rangle $, where $c_{i\tau}$ and $c^\dagger_{i\tau}$ act on unit cell $i$ of sublattice $\tau$, in the ground state of $H$ for different $V_1$ obtained by using the infinite density matrix renormalization group (iDMRG), implemented using the \textsc{tenpy} package \cite{Hauschild2018}, as explained in Ref. [\textcolor{blue}{28}]
. For small interactions $\Delta n$ is close to zero, and grows continuously to one as $V_1$ is increased [\textcolor{blue}{28}]
, signaling a second-order phase transition (see e.g. Ref. [\textcolor{blue}{28}] 
 for a discussion). As shown in Fig.~\ref{ExDFig7}b, we observe a concomitant second nearest-neighbor bond asymmetry, defined as $\Delta t_2= \mathrm{Re}\left[\left\langle c^\dagger_{iA} c_{i+1 A}-c^\dagger_{i B}c_{i+1 B}\right\rangle\right]$, that develops at intermediate values of $V_1$, as expected based on our previous symmetry discussion. As $V_1$ increases, $\Delta t_2$ increases until reaching a maximum, and then decreases as $V_1$ becomes larger, for all cylinder circumferences $L_y=6,8,10$. It is possible to check numerically that setting $t=0$ in Equation \eqref{eq:graphenham}, i.e. in the limit $V_1/t \to \infty$, leads to $\Delta t_2=0$ and $\Delta n=1$, as discussed above.\\

The above results support that a charge-density-wave order with a partial sublattice imbalance, i.e. $0<\Delta n <1$, is {generically accompanied} by a second-nearest-neighbor bond asymmetry, $\Delta t_2\neq 0$, as {argued} in the main text.

{In} the zLL of graphene the wavefunctions at each valley live in different sublattices and thus a full valley polarization implies a full sublattice polarization $\Delta n =1$, in which case $\Delta t_2 = 0$.  When the sublattice polarization is not maximal then a finite $\Delta n$ and $\Delta t_2$ are expected, consistent with what is observed in experiment (see Fig. 5). As mentioned in the main text, this effect can originate {from} Landau level mixing since the sublattice index is not locked to valley index {beyond the zLL} [\textcolor{blue}{31}-\textcolor{blue}{33}]
.


\section{Influence of the Moir\'e superlattice on the CDW phase.}

\begin{figure*}[!b]
	\centering
		\includegraphics[width=1\textwidth]{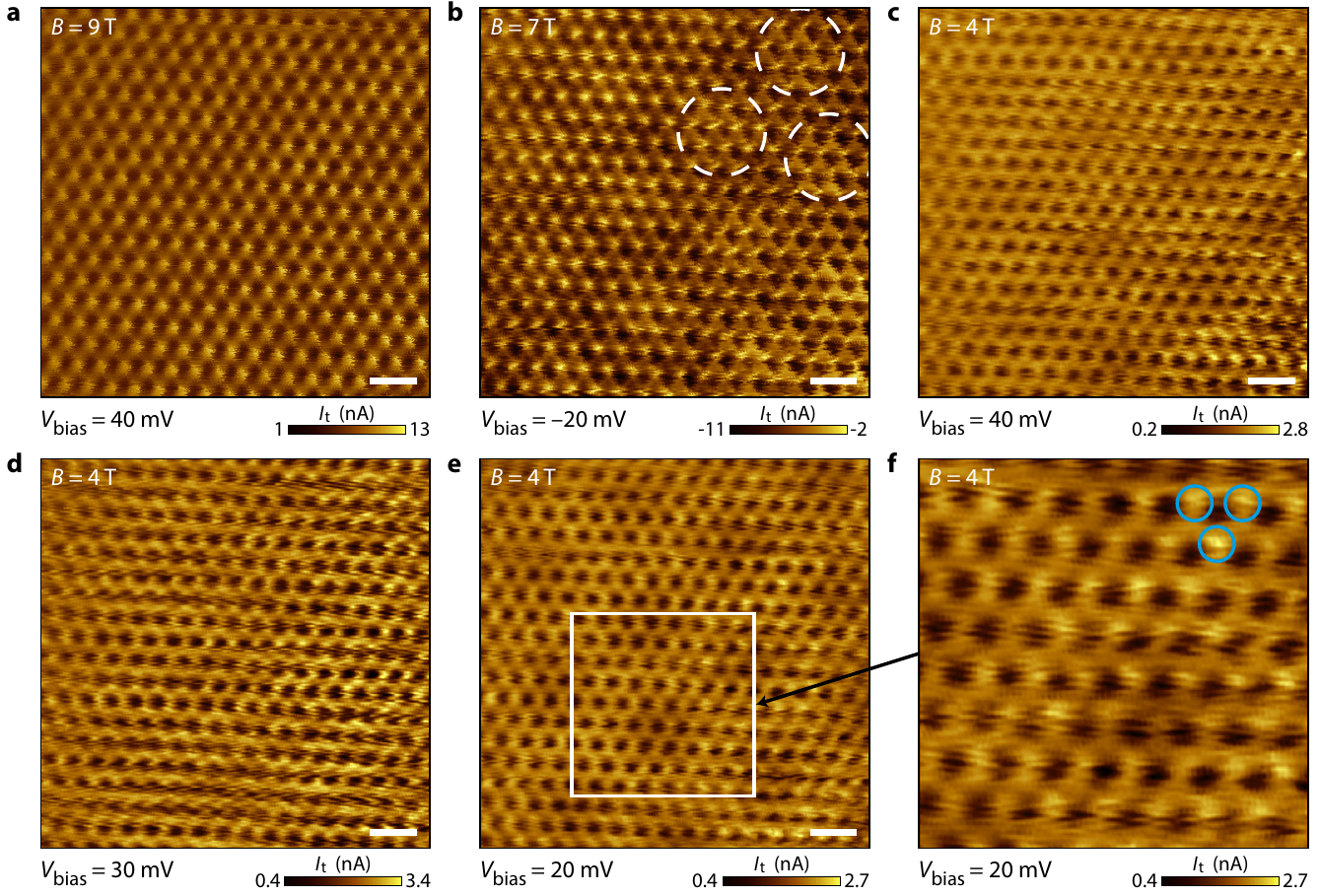}
	\caption{\textbf{Disappearance of the charge-density wave at low magnetic field in sample AC23.} \textbf{a}, CDW at $B=9\:$T. \textbf{b}, CDW at $B=7\:$T. The Moir\'e superlattice is visible but does not perturb the CDW pattern. \textbf{c,d}, Honeycomb lattice with no CDW at $B=4\:$T. \textbf{e}, Honeycomb lattice at $B=4\:$T with residual traces of CDW, see the zoom in \textbf{f} of the white rectangle. Scale bars for all figures are $500\:$pm.}
	\label{ExDFig8}
\end{figure*}

The sample AC23 displays a weak Moir\'e superlattice (weak in the sense that is not always visible in our images). This rises the question of whether the CDW phase we observed was induced by the Moir\'e pattern, which could also break the sublattice symmetry, or not. In such case, we may expect the CDW pattern to rely on that of the Moir\'e, with the sublattice polarization depending on the position inside the Moir\'e superlattice (due to the periodic potential it induces in graphene). Fig. \ref{ExDFig8}b displays a CDW phase observed at $B=7\:$T. The Moir\'e pattern is barely visible but appears as bright spots, such as the ones indicated by dashed white circles. The CDW pattern itself is homogeneous over the whole Moir\'e lattice.\\

\begin{figure*}[ht!]
	\centering
		\includegraphics[width=0.7\textwidth]{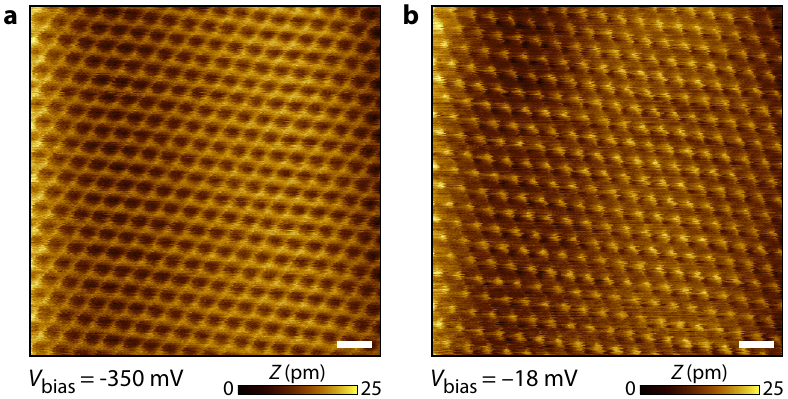}
	\caption{\textbf{Charge-density-wave order in sample AC24.} \textbf{a}, Honeycomb lattice at $B=14\:$T and $V_\text{b}=-350\:$mV observed at $\nu=0$. \textbf{b}, CDW under the same conditions but at $V_\text{b}=-18\:$mV. Scale bars for both figures are $500\:$pm.}
	\label{ExDFig9}
\end{figure*}

STM images of the similar sample AC24 on SrTiO$_3$ did not exhibit any Moir\'e superlattice. Still, in the same conditions, at charge neutrality, we observed signatures of a CDW phase, shown in Fig. \ref{ExDFig9}b, which indicates that the CDW we observe in our hBN/SrTiO$_3$ samples is indeed an intrinsic consequence of many-body interactions at charge neutrality and not due to extrinsic substrate-induced sublattice symmetry breaking.

\section{Disappearance of the CDW order at low magnetic field.}

Figures \ref{ExDFig8}a and b show that the CDW persists at $B=9\:$T and $B=7\:$T on sample AC23. However, Figs. \ref{ExDFig8}c to e at $B=4\:$T does not display the CDW anymore but only the usual sublattice-unpolarized honeycomb lattice.

Finally, it is theoretically expected [\textcolor{blue}{10}] that graphene undergoes a first-order phase transition from the CDW to the F phase. Such a transition should induce the formation of domains in graphene with the coexistence of both phases around the magnetic field at which the transition occurs. When taking a closer look to Fig. \ref{ExDFig8}e taken at $B=4\:$T, see its zoom in Fig. \ref{ExDFig8}f, one can discern some bright dots in some parts of the image (see the blue circles). This residual asymmetry of the honeycomb lattice is reminiscent of the charge-density wave. It is possible that it may constitute a signature of such domains around the transition between the CDW and F phases.

\normalem
\bibliography{MBGS-BIB}
\bibliographystyle{naturemag}